\documentclass[12pt]{article}

\makeatother
\usepackage[utf8]{inputenc}
\usepackage[usenames,dvipsnames]{color}
\usepackage{amsmath}
\usepackage{amssymb}
\usepackage{mathptmx}
\usepackage{authblk}
\usepackage{textcomp}
\usepackage{float}
\usepackage{gensymb}
\usepackage{siunitx}
\usepackage{cancel}
\usepackage[utf8]{inputenc}

\DeclareUnicodeCharacter{02BC}{'}
\usepackage[a4paper, margin=1in]{geometry}

\usepackage[left]{lineno}
\usepackage{mathptmx}
\usepackage[T1]{fontenc}
\usepackage{lineno}
\usepackage{multirow}
\usepackage{array}
\usepackage{rotating}
\usepackage{booktabs}
\usepackage{comment}

\usepackage{url}

\usepackage[utf8]{inputenc}
\usepackage[T1]{fontenc}

\usepackage[numbers,sort,compress,comma]{natbib}
\usepackage[breaklinks,colorlinks,citecolor=black,linkcolor=black,urlcolor=black]{hyperref}

\makeatletter
\AtBeginDocument{%
  \renewcommand\NAT@open{(}%
  \renewcommand\NAT@close{)}%
  \protected@edef\@tempa{\noexpand\def\noexpand\NAT@nmfmt##1{\noexpand\textit{##1}}}%
  \@tempa
}
\makeatother

\usepackage{graphicx}
\usepackage{subcaption}
\usepackage[labelfont=bf]{caption}
\DeclareCaptionLabelSeparator{bar}{ \textbar{} }
\newcommand{\arcsec}{^{\prime\prime}}
\newcommand{\arcmin}{^{\prime}}

\def\jnl@style{\it}
\def\aaref@jnl#1{{\jnl@style#1}}

\newcommand{\apj}{Astrophys. J.}   
\newcommand{\aap}{Astron. Astrophys.}   

\renewcommand{\figurename}{Fig.}

\usepackage{enumitem}

\usepackage[normalem]{ulem}

\usepackage[dvipsnames]{xcolor}
\definecolor{mycolor}{rgb}{0.1, 0.2, 0.3}

\usepackage{authblk}

\usepackage{orcidlink}

\usepackage{fancyhdr}

\title{\textbf{Evidence for the gravity-driven and magnetically-regularized gas flows feeding the massive protostellar cluster in Cep A}}

\author[1$\dagger$]{Panigrahy Sandhyarani \orcidlink{0009-0007-6357-6874}}
\affil[1]{\small Department of Physics, Indian Institute of Science Education and Research Tirupati, Yerpedu, Tirupati - 517619, Andhra Pradesh, India \url{sandhyaranipanigrahy@students.iisertirupati.ac.in}}

\author[1,2$^{\ast\dagger}$]{Chakali Eswaraiah \orcidlink{0000-0003-4761-6139}}
\affil[2]{Department of Physical Sciences, Indian Institute of Science Education and Research (IISER) Mohali,
Knowledge City, Sector 81, SAS Nagar 140306, Punjab, India}
\affil[$\dagger$]{\textit{The two authors contributed equally to this paper and share the first authorship.}}
\author[3]{Di Li \orcidlink{0000-0003-1674-0110}}
\affil[$\ast$]{\small Corresponding author: \texttt{eswaraiahc@iisermohali.ac.in}}

\affil[3]{\small New Cornerstone Science Laboratory, Department of Astronomy, Tsinghua University, Beijing 100084, China \url{dili@tsinghua.edu.cn}}

\author[4]{Enrique V\'{a}zquez-Semadeni \orcidlink{0000-0002-1424-3543}}
\affil[4]{\small Instituto de Radioastronom\'{i}a y Astrof\'{i}sica, Universidad Nacional Aut\'{o}noma de M\'{e}xico, 58090 Morelia, Michoac\'{a}n, Mexico}

\author[4]{Gilberto C. G\'{o}mez \orcidlink{0000-0003-4714-0636}}

\author[5]{Travis J. Thieme \orcidlink{0000-0003-0334-1583}}
\affil[5]{\small Academia Sinica Institute of Astronomy \& Astrophysics, 11F of Astronomy-Mathematics Building, AS/NTU, No.1, Section 4, Roosevelt Road, Taipei 10617, Taiwan}

\author[6]{Manash R. Samal \orcidlink{0000-0002-9431-6297}}
\affil[6]{\small Physical Research Laboratory (PRL), Navrangpura, Ahmedabad 380 009, Gujarat, India}

\author[7]{Jia-Wei Wang \orcidlink{0000-0002-6668-974X}}
\affil[7]{East Asian Observatory, 660 N. A\'{o}h\={o}k\={u} Place, University Park, Hilo, HI 96720, USA}

\author[8]{Shih-Ping Lai \orcidlink{0000-0001-5522-486X}}
\affil[8]{\small Institute of Astronomy and Department of Physics, National Tsing Hua University, Hsinchu 30013, Taiwan}

\author[9]{Wen-Ping Chen \orcidlink{0000-0003-0262-272X}}
\affil[9]{\small Institute of Astronomy, National Central University, Taoyuan 32001, Taiwan}

\author[10]{D. K. Ojha \orcidlink{0000-0001-9312-3816}}
\affil[10]{\small Department of Astronomy and Astrophysics, Tata Institute of Fundamental Research, Mumbai 400005, India}

\date{}

\begin{document}

\maketitle

\textbf{The hierarchical interplay among gravity, magnetic fields, and turbulence in forming massive protostellar clusters remains elusive. We present high-resolution ($\sim$14 arcsec $\simeq$ 0.05 pc) 850 \SI{}{\micro\meter} dust polarization and C$^{18}$O line observations of Cepheus A using JCMT SCUBA-2/POL-2 and HARP. Our analysis reveals aligned gravitational (G), magnetic (B), and velocity fields (K) , with an energy hierarchy of $E_{\mathrm{G}}$ > $E_{\mathrm{B}}$ > $E_{\mathrm{K}}$. Gravity, as the primary driver, induces gas flows and drags in B-field lines. Magnetic tension, as a secondary force, regulates turbulence, enabling ordered flows with an accretion rate of $\sim$ 2.1 $\pm$ 0.4 $\times$ 10$^{-4}$ M$_\odot$ yr$^{-1}$. This challenges the conventional view of B-fields resisting collapse in the clump/hub scale, instead showing cooperation with gravity. The $\sim$0.6 pc clump-scale B-field (with mean PA $\sim$ 45°) aligns coherently with fields at cloud ($\sim$5 pc), core ($\sim$0.05 pc), and disk ($\sim$2000 AU) scales, offering new insights into the role of magnetic fields in multiscale star formation dynamics.}



Teaser: Gravity gathers while magnetic fields guide the gas inflow, together they nurture massive star formation in Cep A.

\section*{Introduction}
Stars are the fundamental building blocks of the Universe, and among them massive stars are of great importance owing to their role in the evolution of interstellar medium and  
affecting star formation and galactic evolution. Massive stars (with mass M > 8 M$_\odot$) predominantly form in the clustered environments ({\it \citealp{lada2003embedded}}), often within dense molecular cloud regions connected to filamentary structures—collectively termed hub-filament systems (HFSs) ({\it \citealp{kumar2020unifying, zhou2022atoms, xu2023atoms}}). In this framework, filaments act as conduits, channeling material from the diffuse cloud into central hubs, which serve as dense reservoirs fueling cluster and massive star formation. Magnetic fields (B-fields) play a crucial role in shaping and regulating these structures ({\it \citealp{ade2016planck,hennebelle2019role}}). Observations of B-fields in large-scale interstellar filaments using optical, near-infrared, and Planck low-resolution sub-millimeter polarization data confined to the low-density regions reveal consistent trends: in diffuse regions, B-fields align with filaments, while in denser regions, they turn perpendicular ({\it \citealp{collaboration2020planck,soler2017relation}}). 


Recent polarization observations indicate that the general trend between filaments and B-fields can vary, particularly with changes in column density along or across filaments connected to hubs. This suggests that the interplay between B-fields and gravity may evolve locally, depending on which force is dominant in a given region. For example, a perpendicular configuration is expected when strong magnetic fields guide gas contraction or inflow onto the filament. U-shaped B-fields may appear when gravity begins to distort the field lines along the filament, while a parallel alignment suggests that B-fields are being fully dragged by gravitational forces ({\it \citealp{pillai2020magnetized,arzoumanian2021dust,wang2024filamentary}}). Such variations are likely to be more pronounced near massive HFSs, where the gravitational field is especially strong ({\it \citealp{wang2020,chung2022}}).


The hourglass-shaped B-field morphology observed in low-mass star-forming cores ({\it \citealp{girart2006,stephens2013,kwon2019}}) is widely regarded as  compelling evidence for the dynamic role of B-fields in regulating gravitational collapse. In the early, magnetically subcritical phase of a molecular cloud core, magnetic tension resists gravitational collapse in the direction perpendicular to the field lines (major axis). However, gas can more freely accumulate or contract along the magnetic field lines (minor axis), enabling contraction in that direction. As the mass increases more rapidly than the B-field strength, the core eventually enters a supercritical state where gravity becomes dominant. During this stage, ambipolar diffusion~--~where neutral particles decouple from ions that are tied to the B-field lines~--~facilitates collapse by drawing field lines inward, resulting in the characteristic hourglass shape ({\it \citealp{shu1987, mouschovias1991}}). 

Such a striking hour-glass configuration has also been observed in regions where massive stars form ({\it \citealp{girart2009, qiu2014}}), including large, complex clouds structured in HFSs such as the Orion and Serpens molecular clouds ({\it \citealp{pattle2017jcmt, sugitani2010near}}). Figure 6 of Beuther et al. ({\it \citealp{beuther2025}}) presents a clear representation of the magnetic field morphology discussed in this context. 
However, in massive star-forming clumps~--~particularly those embedded within high density portions of HFSs~--~the subsequent evolution of these hourglass-shaped B-fields remains poorly understood. Specifically, it is still unclear how B-fields are reconfigured in gravitationally dominant clumps, how they interact with other forces such as gravity and turbulence, and how they govern the formation of massive stars.

Despite extensive studies of HFS using dust continuum and spectral line data ({\it \citealp{wang2019jcmt,wang2020formation,beuther2020gravity}}), the structure and role of B-fields in dense hubs or clumps remain unexplored. This is primarily because the physical conditions within these clumps evolve as massive star formation progresses, making it challenging to trace their pristine morphology of B-field. For instance, in Orion A’s BN/KL region, an explosive event significantly altered the B- field on $\sim$ 5000 AU scales ({\it \citealp{cortes2021explosion}}), yet at larger clump scales, the field remains ordered, as seen in JCMT SCUBA-2/POL-2 observations ({\it \citealp{pattle2017jcmt}}). Similarly, in Serpens Core, protostellar outflows have disrupted the B-field and cloud structure ({\it \citealp{sugitani2010near,kwon2022b}}), causing multiple transitions in the relative orientation between the B-field and dust structures. These examples illustrate how stellar feedback can reshape cloud structures, complicating our understanding of the fundamental interactions between gravity, turbulence, and B-field. However, some regions, such as Cep A, serve as natural laboratories where imprints of these interactions are preserved, offering valuable insights into the fundamental physics governing star formation.

In massive star-forming regions, gravity is typically the dominant force ({\it \citealp{liu2022atoms,sanhueza2021gravity,cortes2024magmar}}), but these are also magnetically predominant with strengths of a few mG (0.1 -- 6 mG) at clump and core scale ({\it \citealp{lai2001phd,curran2007mnras,beuther2023density}}) to a few 100s of mG at the disc scale ({\it \citealp{vlemmings2006aap}}). While gravity primarily drives cloud collapse, the role of B-fields—whether they actively regulate gravitational contraction or are passively shaped by it—remains an open question. Additionally, gravitational collapse generates turbulence ({\it \citealp{li2017taichi}}), and since turbulence  and B-fields are coupled, their combined effects depend on which factor dominates. Investigating the local morphological variations of B-fields, gravity, and turbulence at clump scales is crucial for resolving the hierarchical interplay of these forces, yet remains a significant observational challenge.

\section*{Results}
\begin{figure*}
\centering
\includegraphics[width=12 cm, height=8 cm]{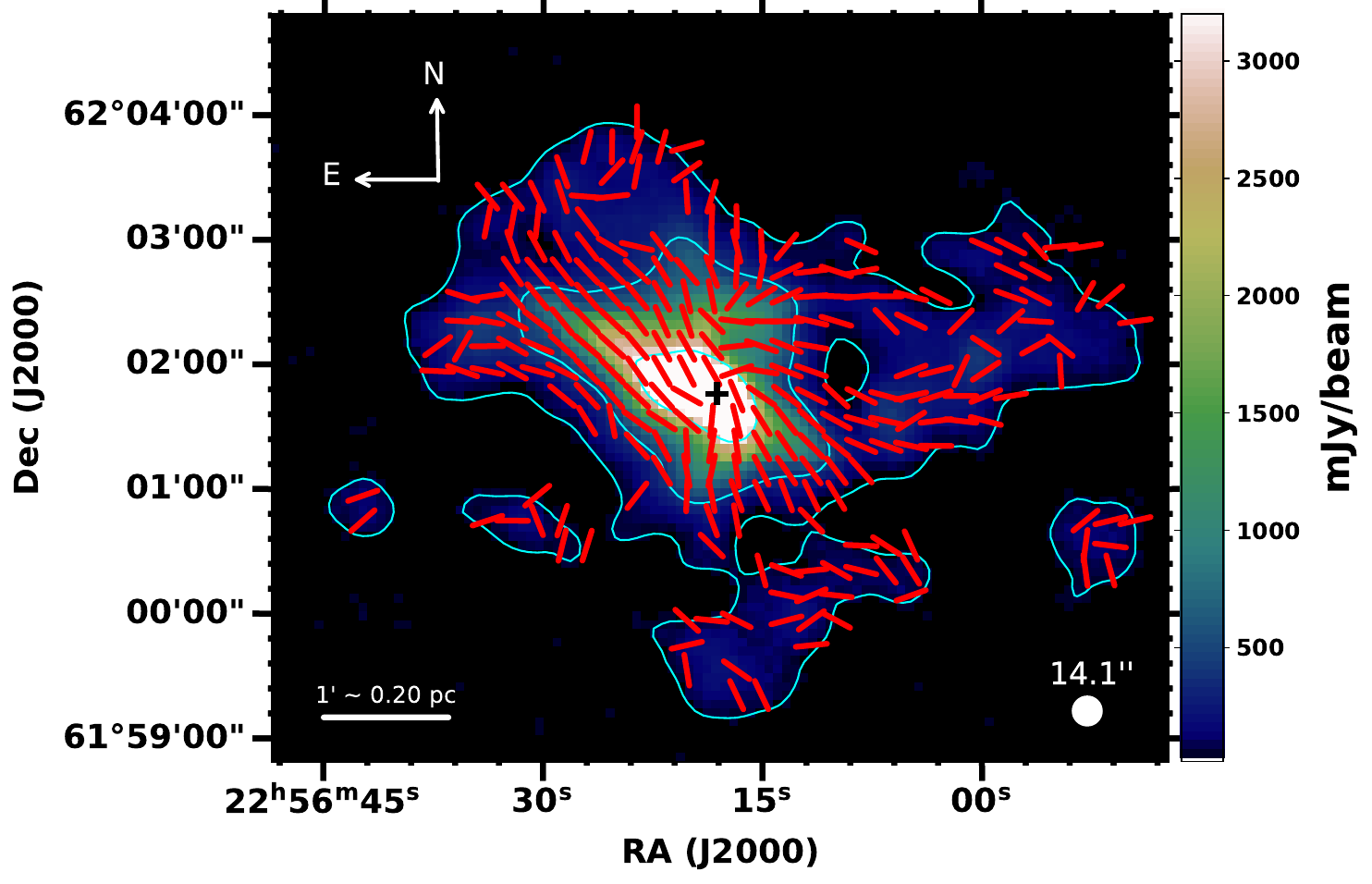}
\caption{{\bf The detailed B-field geometry in the Cep A clump}. The detailed B-field geometry (red segments) of Cep A clump using JCMT SCUBA-2/POL-2 observations. All segments are depicted at equal lengths to visualize the B-field geometry better. The background image is the dust continuum map at 850 \SI{}{\micro\meter}. The cyan contours are drawn at 32, 430, and 3000 mJy/beam. The outermost contour at 32 mJy/beam corresponds to 10$\sigma$, where $\sigma = 3.2$ mJy/beam is the rms noise in the dust continuum map. The color bar on the right denotes the intensity scale across the region. The white circle represents the beam size and the `+' sign represents the position of Cep A HW2 protostar.} 
\label{fig:Bfieldmorphology}
\end{figure*}

In this study, we investigated the second nearest ($\sim$ 725 pc ({\it \citealp{dzib2011vlba}})) massive star-forming clump (1200 M$_{\odot}$ ({\it \citealp{yu1996}})) Cepheus A (Cep A) located in the Cepheus OB3 cloud complex. Cep A hosts an early-type massive protostar (spectral type $\sim$ B0.5, and mass $\sim$ 15M$_{\odot}$) known as Cep A-HW2. The core hosting HW2 is surrounded by a cluster of far-infrared (FIR) sources, indicating the formation of a massive protostellar cluster ({\it \citealp{koppenaal1979}}). The protostar Cep A-HW2 is in an early stage of evolution, characterized by active jets, outflows, and ongoing accretion activity ({\it \citealp{cunningham2009pulsed,sanna2017planar}}). Despite this, it has not yet disrupted the entire parental clump, Cep A, as evidenced by the dust continuum emission map (Figure \ref{fig:Bfieldmorphology}), and thus retains its pristine physical conditions. These characteristics make Cep A an ideal laboratory for examining the detailed structure of B-fields and investigating their role alongside gravity and turbulence in the protostellar cluster environment.

We have performed sensitive sub-millimeter dust polarimetric observations at 850 \SI{}{\micro\meter} towards Cep A using the Submillimetre Common User Bolometer Array (SCUBA-2) in conjunction with the polarimeter (POL-2) on the James Clerk Maxwell Telescope (JCMT) in Hawaii. The B-fields were traced using JCMT but in combination with an old camera, SCUBA, plus polarimeter, POL, (SCUPOL ({\it \citealp{greaves2003submillimetre}})),
however, they were limited in both low- and high-density regions ({\it \citealp{curran2007magnetic,matthews2009legacy}}). Our new observations with POL-2 have delineated the detailed B-field morphology over the scales of $\sim$0.05~--~0.60 pc of Cep A clump as shown in Figure \ref{fig:Bfieldmorphology} (Refer to \hyperref[sec:Methods]{Materials and Methods} for more details). 
The overall B-field structure is quite organized on a larger scale. However, within each of its portions, the orientation of the B-field varies smoothly from large to small scales, consistently pointing toward the densest region of the clump and exhibiting a converging feature towards the center.  

\subsection*{Gas Kinematics and Infalling Structures in Cep A}
Based on our analysis, Cep A emerges as a compelling candidate for HFS. Filamentary features at the cloud scale are clearly visible in the Herschel 250 \si{\micro\meter} map (see left panel of Fig. S\ref{fig:planck_on_herschel_pmo}). To identify gas structures on this scale, we utilized $^{13}$CO (J = 1–0) spectral data from the Purple Mountain Observatory’s Milky Way Imaging Scroll Painting (MWISP) survey and applied the \href{https://fil-finder.readthedocs.io/}{\tt FILFINDER} algorithm. The extracted velocity profiles along the filaments reveal a clear velocity gradient directed from the outskirts toward the central hub (Fig. S\ref{fig:LARGE_SCALE_SMALL_SCALE}).
At the clump scale, we further analyzed C$^{18}$O (J = 3–2) data obtained from the JCMT’s Heterodyne Array Receiver Program (HARP). Velocity gradients derived along four cuts (Fig. S\ref{fig:LARGE_SCALE_SMALL_SCALE}) consistently indicate inward gas motion. These gas kinematics suggest a global convergence of material toward the center, evident at both cloud and clump scales.

To examine the infalling gas structures in Cep A in greater detail, we focused on the HARP C$^{18}$O spectral cube, isolating the velocity range from $-$12.5 to $-$8 km s$^{-1}$—a range free from outflow contamination ({\it \citealp{cunningham2009pulsed}}). Applying the \href{https://dendrograms.readthedocs.io/}{\tt astrodendro} algorithm to the extracted cube reveals a prominent gas structure composed of four extended leaf-like substructures (see Fig. S\ref{fig:dendrogram_structure}).

\subsection*{Interaction Between Outflow and B-Field}
Cep A hosts an active and energetic outflow emanated from the HW2, which is well-traced by shock-excited H$_{\mathrm{2}}$/2.12 \SI{}{\micro\meter} emission (Fig. S\ref{fig:outflows_and_Bfieldmorphology}). 
To investigate the morphological correlation between outflows and the B-field, an offset map is constructed by calculating the difference between the mean position angle of outflows in a given region (determined by the average direction of shock fronts ({\it \citealp{cunningham2009pulsed}}); \hyperref[sec:Methods]{Materials and Methods}) and the B-field. Smaller offset angles, indicated with red pixels, suggest either the outflow has shaped B-fields in the east-west regions or the B-field has guided the outflows. The dominance of one on the other factor can be determined by comparing their energies at different scales. At the disc scale, the Zeeman effect of H$_{\mathrm{2}}$O and OH masers yielded a few 10s to several 100s of mG of B-field strength in the vicinity of HW2 disc ({\it \citealp{vlemmings2006aap}}). These B-fields are dynamically strong enough to launch the outflows from the HW2 protostar ({\it \citealp{pudritz2019role,mignon2023role,oliva2023modeling}}). In addition, B-fields with strengths of few to several mG at the clump scale (present work; see Fig. S\ref{fig:Bfield_lambda}; see also Curran \& Chrysostomou ({\it \citealp{curran2007mnras}})) as well as at core scale ({\it \citealp{beuther2023density}}), could still guide the pc-scale outflows. To strengthen this evidence, we evaluated the outflow energies of the blue- and red-shifted components as (5.73 $\pm$ 0.56) $\times$ 10$^{44}$ and (2.80 $\pm$ 0.30) $\times$ 10$^{44}$ erg, respectively (\hyperref[sec:Methods]{Materials and Methods}). Additionally, we calculated the average magnetic energy ($E_\mathrm{B}$) to be (4.19 $\pm$ 1.35) $\times$ 10$^{46}$ erg, indicating that the B-fields are sufficiently strong to guide the outflows in the east-west direction. We focus on regions unaffected by outflows (North-South region) to analyze the interplay between B-field, gravity, and turbulence to understand their relative importance in the Cep A clump. Note 
that both the magnetic and outflow energy estimates carry uncertainties. This energy comparison is intended to be qualitative, aimed at assessing whether magnetic forces are likely to dominate over outflow-driven feedback in shaping the observed structure. The results suggest that magnetic energy likely exceeds the contribution from outflow feedback at clump scales.


\subsection*{Gravitational vector map}
To examine the significance of the B-field relative to gravity, we produce the global gravitational field vectors at each pixel of the dust emission map. The local gravitational field in a star-forming region is a key factor influencing star formation and evolution. Mass distribution in a star-forming region determines the local gravitational field, which in turn affects the motion and dynamics of the gas and dust in the region. The gravitational field can cause the gas and dust to collapse, form dense cores, and eventually lead to the formation of stars. The local gravitational field can be studied by observing the motion and distribution of gas and dust.

To determine the impact of gravity in the star-forming region, we estimated the projected gravitational vector field using the JCMT 850 \SI{}{\micro\meter} dust continuum map. The direction and relative strength of local gravity at each point in a map can be determined by summing up the dust emission from surrounding pixels. This vector sum, calculated at each pixel, considers the directions and distances of every neighboring pixel, weighted by the dust emission. To use this information to understand the local gravitational pull, it is assumed that the distribution of dust emission accurately reflects the distribution of total mass. This allows for the visualization of the direction of local gravity. The absolute strength of gravity can be scaled using a gas-to-dust mass ratio, but this ratio only impacts the magnitude and not the direction of local gravity vectors.
The local projected gravitational force acting at a pixel position ($\Vec{F_{G,i}}$) can be calculated using the polarization-intensity gradient technique ({\it \citealp{koch2012a,koch2012b}}). This method involves summing up the vectors of gravitational pulls from all neighboring pixel positions, which can be expressed as
\begin{align*}
    \vec {F_\text{G,i}} =  \text K\,I_{\text i} \sum_{\text j=1}^{\text n} \frac{I_{\text j}}{r_{\text {ij}}^2} \hat{r}  
\end{align*}
The equation for the local projected gravitational force at a pixel position ($\Vec {F_{G,i}}$) considers the gravitational constant and the conversion from emission to total column density through the factor $K$. The intensities at pixel positions $i$ and $j$ ($I_\text{i}$ and $I_\text{j}$) and the number of pixels within the relevant gravitational influence ($n$) are also included. The POS projected distance between pixel i and j (r$_{i,j}$) and the corresponding unity vector ($\hat{r}$) are used to determine the local gravitational vector field, which shows the direction and magnitude of the gravitational pull at each selected pixel. For this analysis, only the direction of the local gravitational forces is considered, and the distribution of dust is assumed to be a good approximation for the total mass distribution. The constant $K$ is set to 1, and a lower threshold is introduced to exclude weak and symmetrical diffuse emission from the calculation. Diffuse emissions at distances outside the extension of our maps can be safely discarded. This is because any gravitational force will be completely canceled out by the emission's down-weighting caused by a rapidly decreasing 1/$r^2$ factor and the tendency of the diffuse emission to become more azimuthally symmetrical at greater distances. This framework allows for interpreting and analyzing the local role of gravity and B-fields, their relative importance, spatial variations, systematic features, and statistical properties. In the analysis, we considered pixels within 4$^{'}$ diameter area with intensity higher than 16 mJy beam$^{-1}$ (5$\sigma$), and those outside 4$^{'}$ diameter are masked. The resulting local gravitational vector field at pixel positions where the B-field segment is present is shown in Figure \ref{fig:sinw}(A). It can be seen that the projected gravitational field vectors are directed toward the massive central hub, which indicates that the hub exerts a dominant effect on the overall gravitational field in the area. It is to be noted that this does not capture line-of-sight variations or complex 3D structures. To understand the relative orientation of B-fields and gravity, we produced the absolute offset angle map by taking the difference between the position angles of the gravitational field vector and the B-field position angle (PA) (shown in Figure \ref{fig:sinw}(B)). The smaller offset angles in the North (N) and South (S) regions marked as red pixels reveal that the gravity and B-field are well-aligned with each other. 

\subsection*{Dominance of Gravity and Magnetic Fields in Cep A: Quantitative Analysis}
To quantitatively assess which factor, among gravity and the B-field, is dominant in Cep A, we have employed the term `$\sin\omega$' through the Magnetohydrodynamic (MHD) force equation ({\it \citealp{Koch2018}}) (\hyperref[sec:Methods]{Materials and Methods}). Here, $\omega$ is the offset angle between gravitational field vector and B-field PA shown (Figure \ref{fig:sinw}(b)). 
For the local collapse to occur, the gravitational force must exceed the magnetic force at a specific location, causing the B-field line to bend in response to gravity. This scenario results in a close alignment between the B-field and gravity and hence smaller $\sin\omega$ values, which is true in the N and S regions because a majority of the blue pixels were found (Figure \ref{fig:sinw}(C)). Therefore, $\sin\omega$ analysis hints at how effectively gravity shapes and aligns the field lines in the N and S regions. 

To quantify the local influence of B-field relative to other forces at play, such as gravity and any other pressure gradient (resultant of thermal/kinetic and non-thermal/turbulent gas pressures), we calculated the parameter $\Sigma_B$ ({\it \citealp{koch2012magnetic,Koch2018}}) (Figure \ref{fig:sinw}(D)). This parameter is equivalent to $F_B$/($F_G$ + $F_P$), where $F_B$ is the magnetic force, $F_G$ is the gravitational force, and $F_P$ is the pressure force. This force ratio is calculated by taking the ratio $\sin\psi/\cos\alpha$, where $\psi$ represents the offset angle between the gravitational field vector and the intensity gradient, and $\alpha$ is the angle between the B-field and the intensity gradient. In the N and S region, where the value of $\sin\omega$ is lower, the $\Sigma_B$ values are also less than 1, indicating that gravity dominates over the magnetic field support, pulling the B-fields toward the gravitational trough. The consequence of this scenario would be that 
gravity would exert a maximum acceleration on the gas, inducing and directing the gas flow (as evident from Fig. S\ref{fig:LARGE_SCALE_SMALL_SCALE}) along the gravitational pull. Under the assumption of flux-freezing ({\it \citealp{caselli1998ionization,blackman2013deriving,das2024magnetic}}), this process also drags B-field lines coupled with the gas flows.
%
We calculated the B-field strength using both the Davis-Chandrasekhar-Fermi (DCF) method ({\it \citealp{davis1951strength,chandrasekhar1953problems}}) with various correction factors and the recently introduced Skalidis-Tassis (ST) method ({\it \citealp{skalidis2021high}}). Since the DCF method tends to overestimate the B-field strength, we opted to use the ST method in our analysis, as it incorporates the incompressible modes of turbulence ({\it \citealp{pattle2022magnetic,camacho2023kinetic}}) (for more details, refer \hyperref[sec:Methods]{Materials and Methods}). To elucidate further, we also estimated the cumulative gravitational ($E_\mathrm{G}$), magnetic ($E_\mathrm{B}$), and turbulent kinetic ($E_\mathrm{K}$) energies in the clump as a function of radial distance from the center (Figure \ref{fig:energies}, \hyperref[sec:Methods]{Materials and Methods}). Our analysis shows that $E_{G}$ is predominant throughout the clump, with an average energy of (11.00 $\pm$ 0.03) $\times$ 10$^{46}$ erg. Interestingly, the dominance of $E_\mathrm{B}$ over $E_\mathrm{K}$ (as seen in Figure 3) suggests that the higher B-field strength may suppress turbulence 
and inhibit secondary fragmentation along the conveyor-like flow. This is because the magnetic tension tends to dampen the dynamical instabilities and hence reduces the turbulence along the converging flows ({\it \citealp{heitsch2008fragmentation,heitsch2009effects}}), which makes the flow more ordered.
In addition, while gas moves more freely along the strong field lines, their motion will be hindered in the direction perpendicular to the field lines ({\it \citealp{inoue2008two,inoue2009two,vazquez2011molecular,inoue2012formation,heitsch2014accretion,lazarian2014reconnection,kortgen2015impact,zamora2018magnetic}}). 
Thus, the B-fields can be envisioned as channels directing water to a lake, providing a pathway for the inflowing material. 
\textit{Indeed, gravity and B-field act hand-in-hand such that gravity induces gas flows and also pulls field lines, while the B-field regularizes these flows along the field lines towards the trough, aiding gravity's effort.}
\subsection*{Multiscale B-Field structure and Its Influence on Gas Dynamics}
We strengthen our findings of 
magnetically-regularized gas flows along the North-East (NE) and South-West (SW) orientation based on the comparison of B-fields at various scales ~--~ probed by {\it Planck} in the cloud, JCMT SCUBA-2/POL-2 in the clump (present work), SMA (Submillimeter Array) in the core ({\it \citealp{beuther2023density}}) and MERLIN (Multi-Element Radio Linked Interferometer Network) in the vicinity of circumstellar disc ({\it \citealp{vlemmings2010}}) as shown in Figure \ref{fig:Merged_allscales}. At the cloud scale, the B-field appears uniformly oriented nearly along the SouthEast-NorthWest (SE-NW) orientation with $108\pm 14 \degree$ from North (Figure \ref{fig:Merged_allscales}(A)). However, close to the NE portion of the cloud, they tend to be dragged towards the hub with a position angle of $39.2 \pm 13.6 \degree$ along the filament F1 towards the hub (Figure \ref{fig:Merged_allscales}(A); see also right panel of Fig. S\ref{fig:planck_on_herschel_pmo}). A clear velocity gradient along the filament F1 towards the hub is also evident (Second row in Fig. S\ref{fig:LARGE_SCALE_SMALL_SCALE}). At the clump scale, B-fields appear organized; nonetheless, they are dragged towards the center, thereby aligned along the direction of gas structures, and eventually towards the center, B-fields exhibit a twisted pattern (Figure \ref{fig:Merged_allscales}(B)). At the $\sim$0.005 pc core scale, the B-field is aligned with the south-west elongated structure, and they are connected smoothly to the core (Figure \ref{fig:Merged_allscales}(C)). A comparison of POL-2 segments with those of SMA reveals a similarity between core and clump scale B-field.
At the vicinity of the circumstellar disc, the B-fields, probed using 6.7 GHz methanol maser polarization data,
are oriented along NorthEast-SouthWest (NE-SW) with a mean position angle of $33.3 \pm 17.7 \degree$, which is parallel to the radio jet and perpendicular to the disc. This component of the B-field is coherent with the B-field at the clump and core scales. A toroidal B-field component is also evident as per Figure \ref{fig:Merged_allscales}(D), where the B-field, initially oriented in the NE-SW, tends to be twisted along the disc. 
Vlemmings et al. ({\it \citealp{vlemmings2010}}), based on the relation between
gas density and B-field strength show that gas collapses along the magnetic field lines, suggesting that the B-field likely regularizes accretion onto the disk around the massive protostar Cep A HW2.

The histograms of B-fields across all scales, presented in Figure \ref{fig:Bfield_multiscale}, evidence the dominant B-field component centered around 45$\degree$ with orientations spanning from 150$\degree$ to 180$\degree$ or 0$\degree$, and to 60$\degree$. Therefore, gravity-driven and magnetically-regularized collapse in the N and S regions is evident at the clump-to-disc scale, and such an important finding is further supported by the signature of bend in the cloud-scale B-field and a similarity in the morphologies of core- and disc-scale B-fields. The observed alignment of the B-field with the jet on small scales does not contradict the gravitational dragging observed at larger scales. Instead, it illustrates a coherent evolutionary picture, where the role of clump-scale B-fields shifts from being passively bent by gravity during infall to actively aligned along the jet axis of the protostar Cep A HW2.
\subsection*{Comparison of observation to the Cassen-Moosman-Ulrich (CMU) model}
To concrete our findings regarding whether the gas structures collapse under gravity, \textit {firstly}, we compared the data points of the dendrogram structures in the position-position-velocity (PPV) space with the data points in the trajectories produced by the Cassen-Moosman-Ulrich (CMU) model ({\it \citealp{ulrich1976infall,cassen1981formation,chevalier1983enviroments}}) (Figure \ref{fig:cmu_ppv}; see \hyperref[sec:Methods]{Materials and Methods} for details on the CMU model). The comparison between the CMU modeled trajectories and the observed dendrogram data in the PPV space shows a 76\% match, indicating it is likely a gravitationally collapsing structure. The three-dimensional gas structure in PPV space is projected into two-dimensional position-position (PP) and position-velocity (PV) plots (middle and bottom panel of Figure \ref{fig:cmu_ppv}).
As shown in PP space, a close alignment between B-fields and the gas flows represented by CMU-modeled trajectories signifies a scenario where the initial B-fields could have been dragged in by gravity,
 consequently they tend to regulate the gas flows along the direction of gravitational pull towards the  potential trough.
\textit{Secondly}, by comparing the gravity vectors with the infalling gas structures inferred by the CMU model trajectories, as shown in the middle panel of Figure \ref{fig:cmu_ppv}, we conclude that the gas velocity field is also aligned along the gravity. By determining the mass of this infalling dendrogram structure to be 120 $\pm$ 47 M$_{\odot}$, we calculated the envelope infall rate as (2.1 $\pm$ 0.4) $\times 10^{-4}$ M$_{\odot}$ yr$^{-1}$ (\hyperref[sec:Methods]{Materials and Methods}). This infall rate matches with the required accretion rate to overcome the radiation pressure and form massive stars ({\it \citealp{beuther2002, zhang2005,grave2009,hosokawa2009}}). Such high infall rates are crucial for the growth of protostars to emerge as massive stars.
\subsection*{Discussion}
In this study, we investigate the detailed B-field morphology in the massive star-forming clump Cep A, which hosts a young protostellar cluster. Using high-resolution submillimeter dust polarization observations at 850 $\mu$m from JCMT SCUBA-2/POL-2, we map the clump-scale B-field structure. Further C$^{18}$O molecular line data from JCMT/HARP reveal filamentary gas structures that appear to align with the local magnetic field morphology. Although the overall B-field geometry is complex, it remains locally organized, with field lines generally converging toward the clump center. We compare the clump scale B-field structure with those at larger and smaller scales using the data 
from Planck (cloud scale), SMA (core scale), and MERLIN (disc scale). 
At the cloud scale, the B-field is predominantly oriented along SouthEast-NorthWest (SE-NW) direction. However along a prominent filamentary structure, the field lines appear to be drawn inward toward the center of the cloud. This realignment coincides with a pronounced velocity gradient, suggesting magnetically-regulated gas flows along a Northeast–Southwest (NE–SW) direction.
Toward smaller, core scales, the B-field becomes aligned with the gas morphology, displaying twisted patterns near the center. Across all spatial scales—from the diffuse cloud to the disc—we identify a consistent B-field component oriented at approximately 45°, which appears to channel material through a coherent accretion flow, regulating mass transfer from the large-scale cloud reservoir down to the central protostar and disc. Near the protostar HW2, the B-field becomes progressively aligned with the thermal radio jet, which extends out to $\sim$ 2000 AU ({\it \citealp{torrelles1993circumstellar}}). At the smallest spatial scales, observations show that the B-field is parallel to the jet and perpendicular to the disk plane. Rather than contradicting our earlier interpretation, we interpret this as an evolutionary, scale-dependent transition in the B-field morphology.


Our key findings try to answer the two intriguing questions: (1) what is the next stage to the U-shaped or hour-glass B-fields in the star forming regions? (2) what is the exact role of such B-field configuration? 
In the massive clump Cep A, we identify a well-defined energy hierarchy, where gravitational energy ($E_{\mathrm{G}}$) surpasses both magnetic ($E_{\mathrm{B}}$) and turbulent kinetic energy ($E_{\mathrm{K}}$), following the order $E_{\mathrm{G}}$ > $E_{\mathrm{B}}$ > $E_{\mathrm{K}}$. The B-field segments, gravity vectors and turbulent velocity fields are largely aligned, suggesting a coherent dynamical structure in which gas is funneled along B-field lines under the influence of gravity. Gravity acts as the primary driver, pulling gas toward the gravitational potential well. Due to flux freezing, the B-field remains tightly coupled to the infalling gas, resulting in the field lines being dragged inward. While the B-field may have initially displayed a U-shaped or hourglass morphology, the overwhelming gravitational influence reshapes it into a sharper, V-shaped configuration. This transformation provides the first observational evidence of a scenario where B-fields no longer resist collapse but instead align with gravitational field vectors, channeling material inward. As the second dominant force, the B-field plays a vital regulatory role: magnetic tension suppresses dynamical instabilities, mitigates turbulence, and organizes chaotic flows into coherent, converging streams. Gas flows more easily along strong B-field lines and is restricted across them, effectively funneling material in a way similar to water flowing along carved channels into a lake. In this dynamically cooperative framework, gravity drives collapse and stretches field lines, while the B-field guides and regulates the flow, collectively enhancing the efficiency of star formation. These results challenge the conventional paradigm that B-fields resist collapse, revealing instead they can dynamically cooperate with gravity to support and guide the star formation process.

Within this region, matter is funneled inward under the influence of gravity in a magnetically-regulated conveyor belt fashion, progressing from the cloud scale down to the clump, core, and disc. As this process unfolds, a mass-segregated cluster forms, with the most massive star at the center. The proximity of Cep A HW2 allows us to zoom in on the clump and its fragmented sub-structures. Cep A hosts a protostellar cluster within a radius of 1$\arcsec$ (725 AU). This is evidenced based on the observed three young protostars, HW2, HW3c, and HW3d, forming within a projected area of $\approx$ 0.6 $\times$ 0.6$\arcsec$ (400 $\times$ 400 AU$^2$) and three hot cores HC, HC2, and HC3 ({\it \citealp{curiel2001detection,curiel2006large}}). Of these six, HW2 stands out as the most luminous source in Cep A with L$\approx$ 10$^4$ L$_{\odot}$, suggesting a mass equivalent to 15 M$_{\odot}$ ({\it \citealp{cunningham2009pulsed}}). Thus, we propose accretion is likely active in the hub, where the envelope is undergoing gravity-driven magnetically regulated collapse with an infall rate of (2.1 $\pm$ 0.4) $\times 10^{-4}$ M$_{\odot}$ yr$^{-1}$. 

Our study, therefore presents for the first time a new insights into how B-field and turbulent gas flows passively assist the active role of gravity in the hub for the formation of a protostellar cluster. 

\clearpage

\begin{figure*}
    \centering

    \includegraphics[width=15.5 cm, height=11 cm]{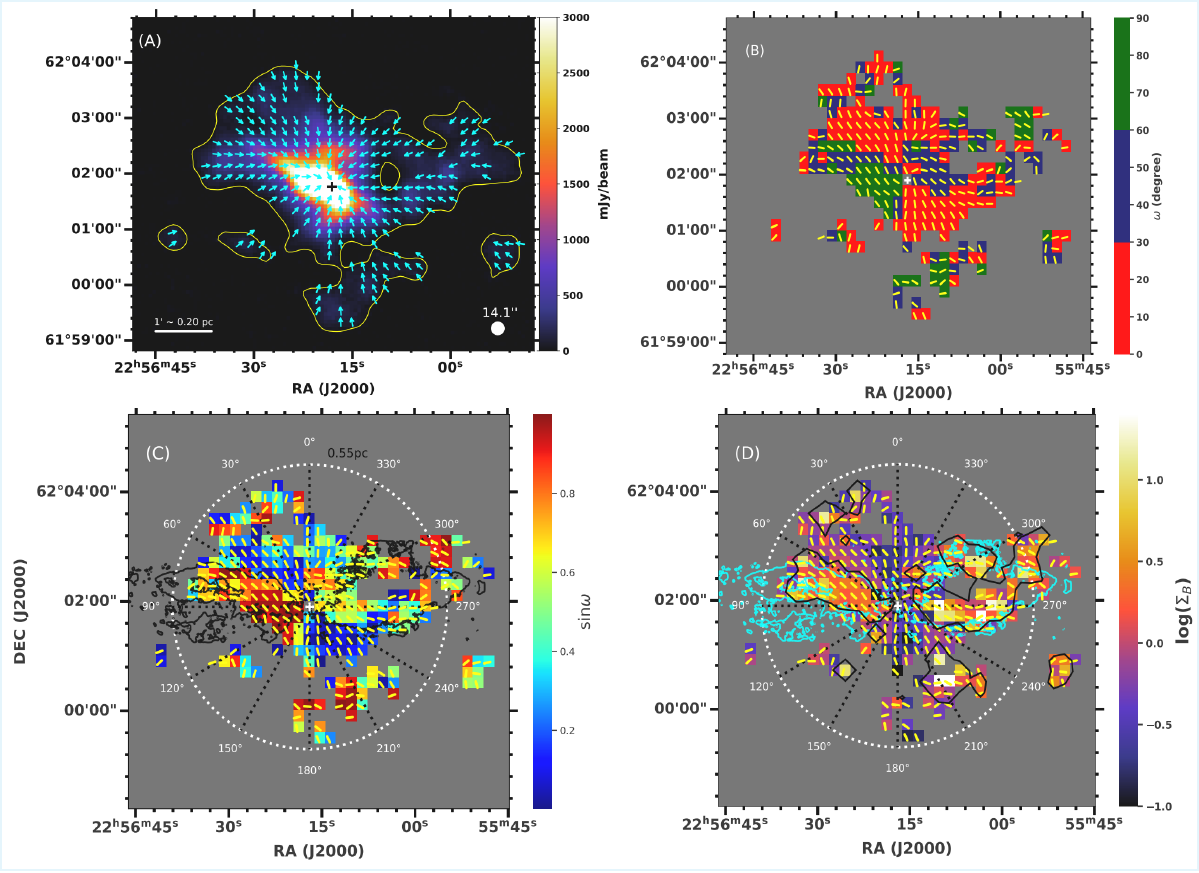}
    
    \caption{{\bf Metrics to understand the role of B-field w.r.t gravity}. ({\bf A}) The gravitational field vector map, represented by cyan vectors,  overlaid on the POL-2 Stokes I map.  Yellow contour corresponds to POL-2 Stokes I of 32 mJy/beam. ({\bf B}) The offset angle ($\omega$) map reveals the morphological correlation between the fields of B-field and gravity. The color bar depicts the color scheme according to different omega values. 
    ({\bf C}) Represents the $\sin\omega$ map. To identify the regions with different $\sin\omega$ values, the entire clump is 
    depicted with an azimuthal angle 0$\degree$~--~360$\degree$ (white dotted circle) and radius (black dotted lines), and ({\bf D}) $\Sigma_{B}$ map in logarithmic scale with base 10. The black contour encloses pixels with $\Sigma_{B}$ > 1. This implies that the pixels outside this contour appear blue with $\Sigma_{B}$ < 1. The yellow segments depict the B-field orientation traced by POL-2. Black and cyan contours in panels ({\bf C}) and ({\bf D}) are the same, depicting the spatial distribution of H$_{\mathrm{2}}$ 2.12 micron shocked emission of 200 mJy/beam. The pixels with smaller $\sin\omega$ and $\Sigma_{B}$ values are evident within the azimuthal angles 330$\degree$ to 30$\degree$ in the north and 150$\degree$ to 240$\degree$ in the south (in the anticlockwise direction). The pixels distributed within these azimuthal angles have the B-field position angles lying between  
    150$\degree$ to 180/0$\degree$, to 60$\degree$. 
    The `+' marks the HW2 protostar in all the panels.}
\label{fig:sinw}
\end{figure*}

\begin{figure*}
\includegraphics[width=14 cm, height=11 cm]{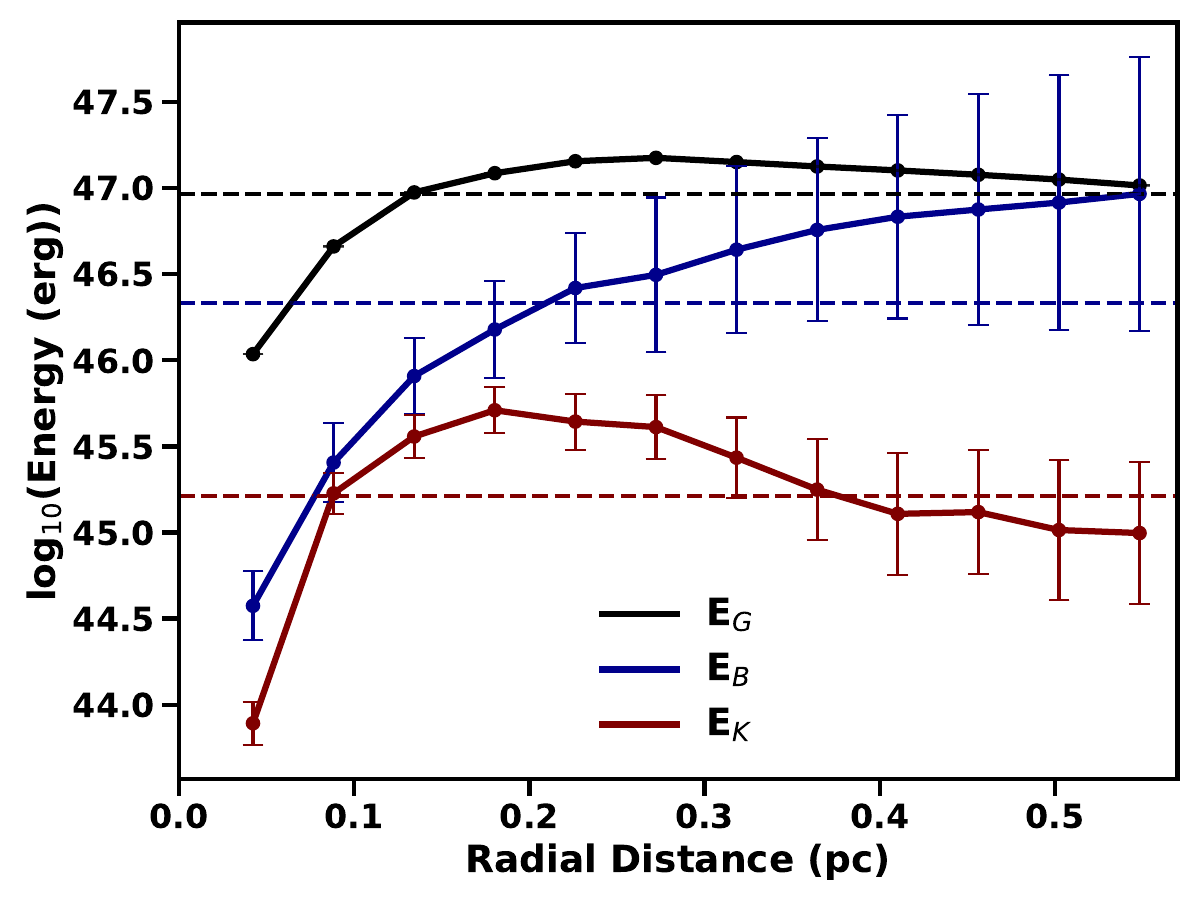}
\caption{{\bf Radial distribution of cumulative energies.} Radial distance versus gravitational ($E_\mathrm{G}$), magnetic ($E_\mathrm{B}$), and kinetic ($E_\mathrm{K}$) energies. Horizontal dashed lines indicate the average energy values. Uncertainties in $E_\mathrm{G}$ are small and comparable to the size of the symbols.}
\label{fig:energies}
\end{figure*}


\begin{figure*}
\centering
\includegraphics[width=16cm, height=12cm]{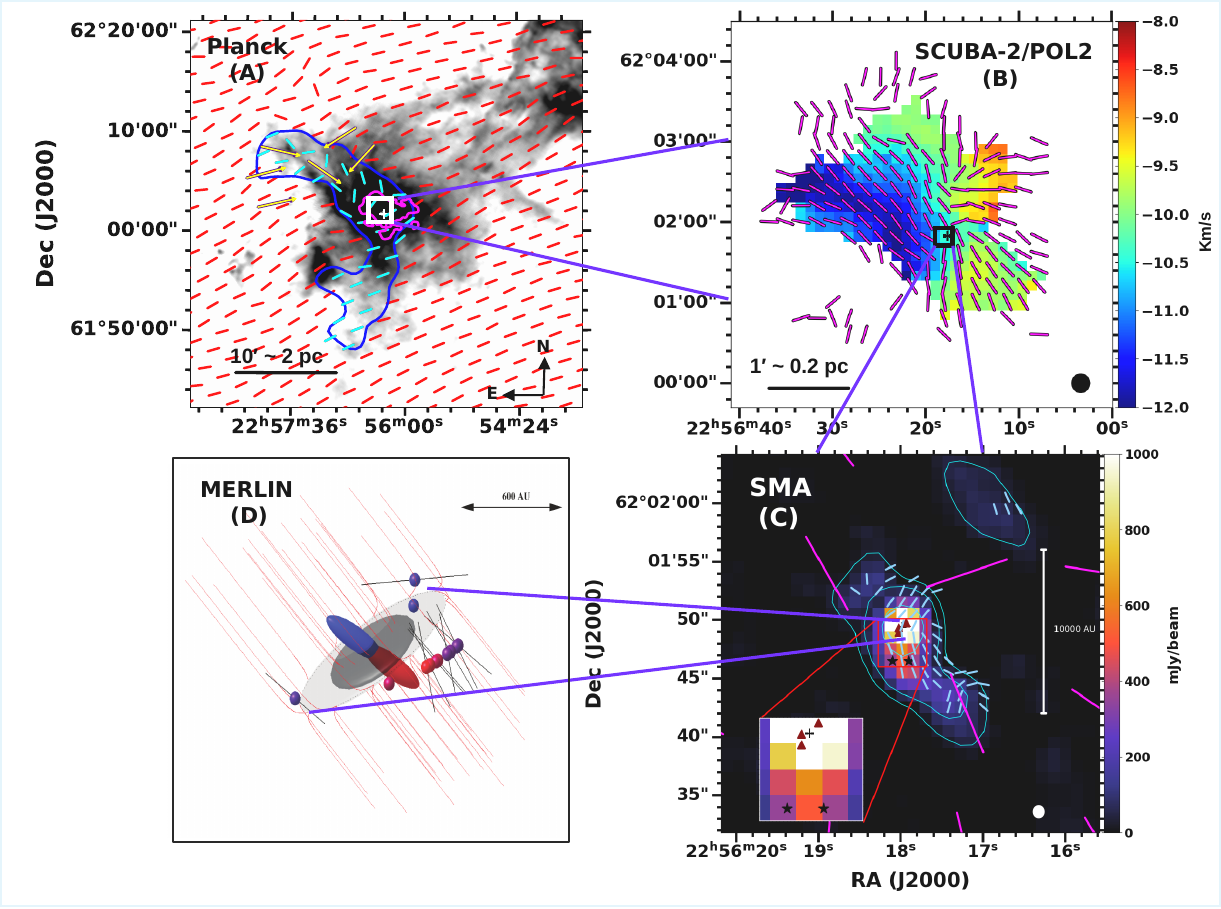}
\caption{{\bf Multiscale B-field structure}. Evolution of B-field structures in the Cep A region from large-scale cloud to small-scale clump, core, and disc scale of the massive protostar Cep A HW2. ({\bf A}) {\it Planck} 850 \SI{}{\micro\meter} B-field segments (resolution $\sim$ 5$\arcmin$ and pixel size 1$\arcmin$) overlaid on the {\it Herschel}/SPIRE 250 \SI{}{\micro\meter} map using red segments. The blue contour, drawn at 3 K km/s, traces filament F1 based on the $^{13}$CO moment 0 map 
(see Fig. S\ref{fig:LARGE_SCALE_SMALL_SCALE}). The B-field segments lying within that contour are marked in cyan. The possible gravity driven bend in the B-field is shown by yellow arrows. The magenta contour represents the POL-2 dust continuum emission at 32 mJy/beam. ({\bf B}) Clump-scale B-fields based on the POL-2 data are shown with magenta segments. The background image is the velocity map revealing a prominent gas structure identified via dendrogram analyses performed on the HARP C$^{18}$O data.
({\bf C}) Core-scale B-fields based on high resolution ($\sim$ 3$\arcsec$) observations at 875 \SI{}{\micro\meter} ({\it \citealp{beuther2023density}}) using SMA are shown with gray segments. A few magenta segments of POL-2 are also shown for comparison. The plus and star symbols correspond to three protostars HW2 (top), HW3c (right), and HW3d (left), while the triangle symbols represent three hot cores HC (top), HC2 (middle), and HC3 (bottom) ({\it \citealp{comito2007high}}), and ({\bf D}) B-fields in the circumstellar disc of Cep A HW2 traced by methanol maser (colored filled circles) polarization observations using the MERLIN telescope at 6.7 GHz (beam size $\sim$ 30 mas) are shown with black segments. The thin red lines oriented in NE-SW trace the proposed B-field morphology piercing through the circumstellar disc. The panel ({\bf D}) has been adapted from Figure 6 of W.H.T. Vlemmings et al. ({\it \citealp{vlemmings2010}}) `Magnetic field regulated infall on the disc around the massive protostar Cepheus A HW2.’ {\it Monthly Notices of the Royal Astronomical Society} (404) 1 (2010): 134-143. Reproduced by permission of Oxford University Press on behalf of the Royal Astronomical Society.
}
\label{fig:Merged_allscales}
\end{figure*}



\begin{figure*}
\centering
\includegraphics[width=10cm, height=20cm]{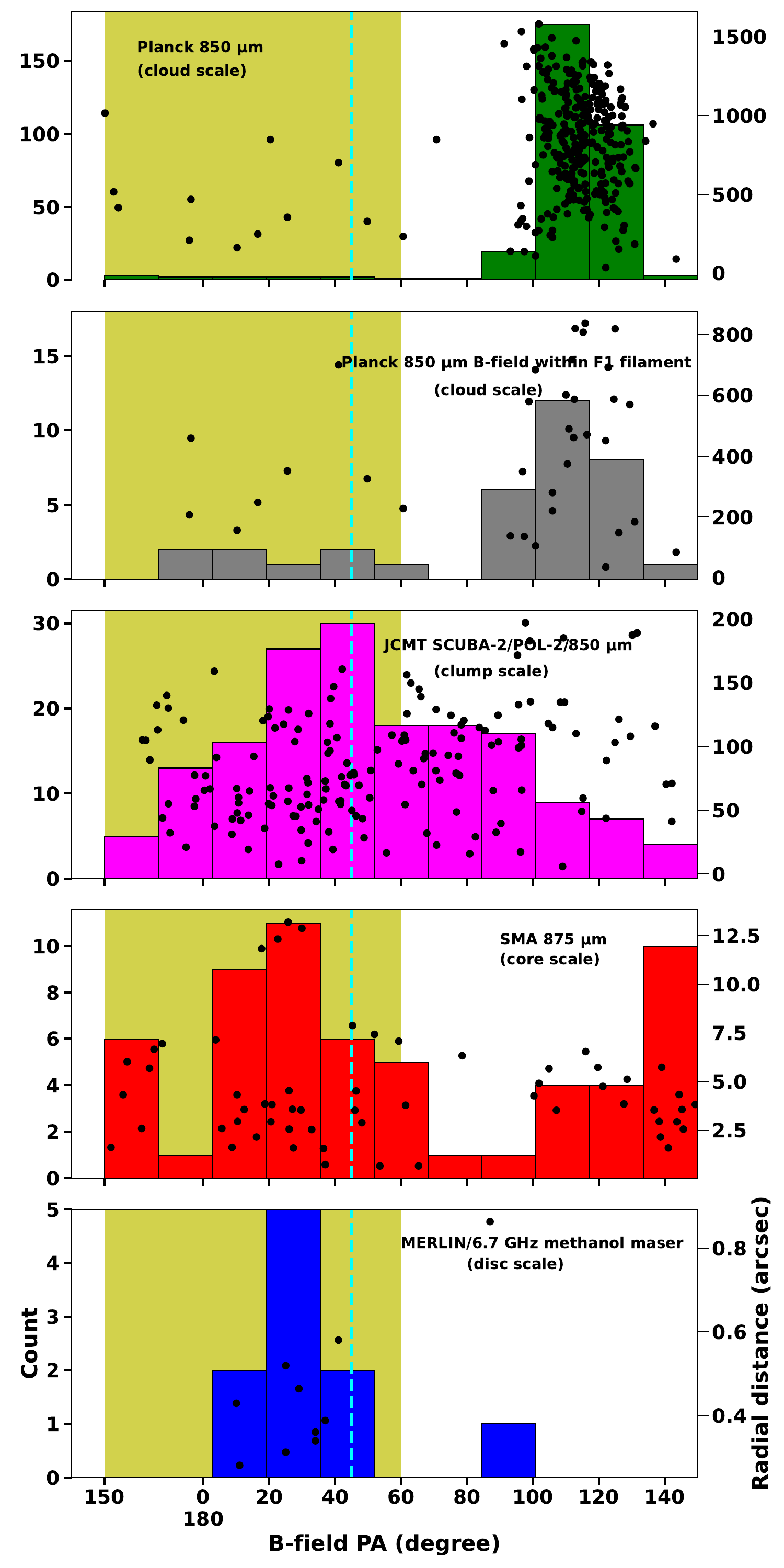}
\caption{{\bf The histograms of the B-field position angles at different scales.} The scatter plot represents the B-field position angles versus radial distance. The cyan dashed line drawn at $\theta$ = 45$\degree$ marks the position angle of the radio jet from HW2. Yellow shaded area denotes the prominent B-field angles, 150$\degree$ to 180$\degree$/0$\degree$, to 60$\degree$, present mainly in N and S regions Cep A (see Figure \ref{fig:sinw} for more details). 
}
\label{fig:Bfield_multiscale}
\end{figure*}

\begin{figure*}
    \centering
\includegraphics[width=15.5cm, height=16cm]{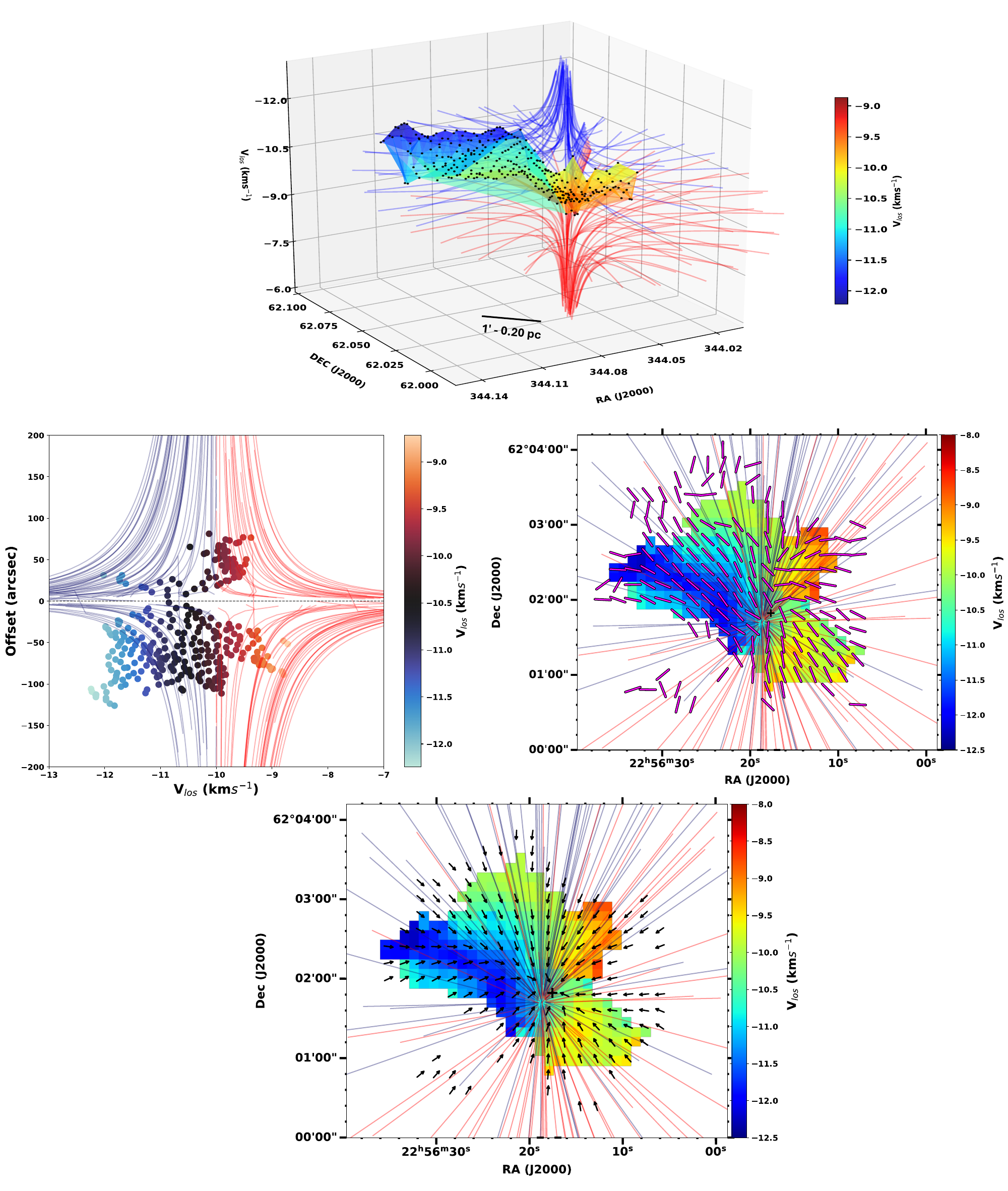}
\caption{{\bf Comparison of CMU modeled trajectories to the observed gas structures}. (Top): Position-Position-Velocity (PPV) data of the gas structures identified based on the dendrogram analyses on JCMT/HARP C$^{18}$O data. The matched trajectories from the CMU model are shown as red and blue-shifted lines. (Middle left): Projected position-velocity (PV) plot showing the matched  CMU modeled trajectories (blue and red lines) to the observations (scattered data). The data points extracted from dendrogram analyses are color-coded according to their velocities. 
(Middle right): Projected position-position (PP) plot showing matched trajectories to the data points. POL-2 B-field segments are also overlaid on them, and (Bottom): Gravitational field vectors (black vectors; same as those shown in Figure \ref{fig:sinw}(a)) and CMU-modeled trajectories are overlaid on the centroid velocity (moment 1) map of the dendrogram structure. The `+’ sign represents the HW2 protostar.}
\label{fig:cmu_ppv}
\end{figure*}
\clearpage
\section*{Materials and Methods}\label{sec:Methods}

\subsection*{Observations and data reduction}
We carried out polarization continuum observation towards massive star-forming region Cep A with the reference position at $\alpha_{\text J2000.0}$=$22^{\text h} 55^{\text m} 51.44^{\text s}$, $\delta_{\text J2000.0}$=$62^\circ 01' 51.3''$
 with SCUBA-2 POL-2 mounted on the JCMT (Project code: M19BP061; PI: Eswaraiah Chakali). SCUBA-2/POL-2  achieved a better  sensitivity ($\sim$ 3.2 mJy/beam) compared to the previous generation instrument SCUBA/POL ($\sim$ 29 mJy/beam) ({\it \citealp{matthews2009legacy}}). All these observations were taken in Band-2 weather conditions with opacity $\tau_{225}$GHz span a range from 0.05 to 0.08. In this study, a single observation using the POL-2 instrument involved a 40-minute observing session employing the POL-2 DAISY scan pattern ({\it \citealp{friberg2016pol}}). JCMT has an effective beam size of 14.1 $\arcsec$ ($\sim$ 0.049 pc) ({\it \citealp{dempsey2013mnras}}). The obtained data were subsequently processed through a two-step reduction procedure utilizing the \href{http://starlink.eao.hawaii.edu/docs/sun258.htx/sun258ss73.html}{\tt pol2map} routine, which was integrated into the Sub-Millimeter User Reduction
Facility \href{http://starlink.eao.hawaii.edu/docs/sun258.htx/sun258.html}{\tt SMURF} package of Starlink software ({\it \citealp{jenness2015automated}}). Continuum polarisation at both 450 and 850 \SI{}{\micro\meter} was seen concurrently. Here in this study, the 850 \SI{}{\micro\meter} data is taken into consideration.

For the clump scale ($\sim$0.6 pc) analysis we have used the $\sim$14$\arcsec$ resolution  $^{13}\mathrm{CO}$ and C$^{18}\mathrm{O}$ ($J = 3\!-\!2$) transition spectral line observations at 330.59\,GHz and 329.33\,GHz, respectively. The observations were carried out using JCMT/HARP as part of the extended observing program (Project Code: M22BP020; PI: Eswaraiah Chakali). These observations were performed in Band 2 weather conditions. The data reduction was carried out using the ORAC Data Reduction (ORAC-DR) pipeline and the Kernel Application Package (KAPPA ({\it \citealp{currie2008starlink}})) within Starlink, as detailed by Buckle et al. ({\it \citealp{buckle2009harp}}).

The $\sim$50$\arcsec$ resolution  $^{13}\mathrm{CO}$ ($J = 1\!-\!0$) molecular line observations at 110.201\,GHz were used for the cloud-scale analysis ($\sim$5 pc). These data were obtained with the 13.7-m radio telescope as part of the Milky Way Imaging Scroll Painting (MWISP) survey, led by the Purple Mountain Observatory (PMO). These observations were done simultaneously using a 3 $\times$ 3 beam sideband-separating Superconducting Spectroscopic Array Receiver (SSAR) system ({\it \citealp{shan2012development}}) and using the position-switch on-the-fly mode, scanning the region at a rate of 50$^{''}$ per second. For details, refer Su et al. ({\it \citealp{su2019milky}}).
The {\it Planck}/353 GHz (resolution $\sim$ 5 $\arcmin$) polarization data, consisting of Stokes I, Q, and U maps, were extracted from the {\it Planck} Public Data Release 2 ({\it \citealp{planck2016planck}}) using the Multiple Frequency Cutout Visualization (PR2 Full Mission Map with PCCS2 Catalog; \url {https://irsa.ipac.caltech.edu/applications/planck/}). The pixel size and beam size were approximately 1$^{'}$ and 5$^{'}$, respectively. We processed the data and obtained the B-field map following the procedures outlined in Section 3.4 of Baug et al. ({\it \citealp{baug2020alma}}) and the references therein.

\subsection*{Correlation between outflow and B-field}
To investigate this further, an offset angle map is produced by taking the difference between the pixel-wise B-field angle and large-scale outflow position angle. The latter was obtained in the eastern region by averaging the angles of ejections from the chains of H$_2$ knots. These include the two eastern components of HH 169 with position angles (PAs) of approximately 80$\degree$ and 65$\degree$, a bright compact H$_2$ bow at around 55$\degree$, and the current orientation of the jet at 45$\degree$. For the western part of the outflow, we used the PA of HH 168 at approximately 100$\degree$ (please refer to Figure 6 of Cunningham et al. ({\it \citealp{cunningham2009pulsed}})).
A one-to-one correlation between the large-scale outflow and the B-field is seen in the East-West region (Fig. S\ref{fig:outflows_and_Bfieldmorphology}). Although the H$_2$ emission spatially overlaps with the blue-shaded region, our combined morphological and spectral analyses support the conclusion that the underlying dust-emitting region traces relatively quiescent material outside of the outflow cavity walls.

\subsection*{Sigificance of $\sin\omega$ map}
The ideal magnetohydrodynamics (MHD) force equation can be written as ({\it \citealp{Koch2018}}),
\begin{equation}
    \rho \: ( \frac{\partial }{\partial t} + v .\nabla)\: v = - \nabla P - \rho \nabla \phi + \frac{B^2}{4\pi R}\: \sin\omega\: g + \frac{B^2}{4\pi R}\: \cos\omega\: n_g,
\end{equation}
where $\nabla \phi$ represents the direction of local gravity and the term $\frac{B^2}{4\pi R}$ $\sin\omega$ can be used to quantify the effectiveness of the B-field in opposing gravitational field. In particular, the factor $\sin\omega$ defines the fraction $\frac{B^2}{4\pi R}$ that can work against gravity to slow down or prohibit the gas motion driven by gravity. Larger offset angles (Figure~\ref{fig:sinw}(b)), as well as larger $\sin\omega$ values (Figure~\ref{fig:sinw}(c)) in the E-W region, suggest nearly orthogonal orientation between gravity and B-field. Since outflows have impacted the E-W region, we do not interpret $\sin\omega$ in terms of gravity versus B-field.

\subsection*{Column density (N$_\mathrm{H_2}$) map}
We constructed the N(H$_2$) and dust temperature ($T_{\text d}$) maps through spectral energy distribution (SED) fitting to all the data points of PACS 160 \SI{}{\micro\meter} (11.4$\arcsec$), SPIRE 250 \SI{}{\micro\meter} (18.1$\arcsec$), SPIRE 350 \SI{}{\micro\meter} (24.9$\arcsec$), SPIRE 500 \SI{}{\micro\meter} (36.4$\arcsec$) and JCMT 850 \SI{}{\micro\meter} (4$\arcsec$) by keeping column density and temperature as free parameters. These are produced using the following relation ({\it \citealp{kauffmann2008mambo}}),
\begin{align}
    S_\nu^{\text{beam}} = & \left(\frac{N_{H_2}}{2.02 \times 10^{20} \ \text{cm}^{-2}}\right) 
    \times \left(\frac{\kappa_{\nu}}{0.01 \ \text{cm}^2 \ \text{g}^{-1}}\right) 
    \times \left(\frac{\Theta_{\text{HPBW}}}{10 \ \text{arcsec}}\right)^2 \nonumber\\
    & \times \left(\frac{\lambda}{\text{mm}}\right)^{-3} 
    \times \left(\exp\left(1.439 \left(\frac{\lambda}{\text{mm}}\right)^{-1} \left(\frac{T_\text{d}}{10 \ \text{K}}\right)^{-1}\right) - 1 \right)^{-1}.
\end{align}
Eventually, we produced the 4$\arcsec$ N(H$_2$) map using the 4$\arcsec$ pixel size POL-2 850 \SI{}{\micro\meter} continuum emission with the low resolution (36.4$\arcsec$) dust temperature map produced using the above equation. Here, we assume that within 36.4$\arcsec$ beam size, the 4$\arcsec$ size pixels with different intensities have the same dust temperature. For more details, refer to Eswaraiah et al. ({\it \citealp{eswaraiah2020unveiling}}) and the references therein. 

\subsection*{B-field strength using Davis-Chandrasekhar-Fermi (DCF) method}
The DCF method has been used to estimate the B-field strengths in Cep A \citep{davis1951strength,chandrasekhar1953problems}. This method assumes that the perturbations in the B-field are Alfv{\'e}nic, which means the deviation in angle from the mean field direction is due to the distortion by small-scale non-thermal motions. Using this method, we can generate a map of the B-field strength by utilizing three quantities: polarization angle dispersion, gas number density, and non-thermal velocity dispersion. The moving box approximation method ({\it \citealp{hwang2021jcmt}}) has been adopted to produce the B-field angle dispersion map.
The number density map is obtained by dividing the N$_\mathrm{H_2}$ map by the witdh of the clump ($\sim$ 0.32 pc). Since the pixel and beam sizes of number density and velocity dispersion maps (4$\arcsec$, and 7.27$\arcsec$, respectively) differ from those of the angle dispersion map (12$\arcsec$), we regridded these maps to a common bin size of 12$\arcsec$, using the Kernel Application Package (KAPPA ({\it \citealp{currie2008starlink}})) task \texttt{wcsalign} and applied gaussian convolution to achieve a common beam size of 14$\arcsec$.
Then, we inserted the values of these three quantities and obtained the plane-of-sky B-field strength (B$_\text{pos}$) by the equation:
\begin{align}
    B_{\text{pos}}[\mu\text{G}]=Q \sqrt{4\pi \rho} \frac{\sigma_v}{\sigma_\theta} \approx 9.3 \sqrt{n(\text{H}_2)[\text{cm}^{-3}]} \frac{{\Delta}V[\text{km s}^{-1}]}{\sigma_\theta [\text{degree}]}, 
    \label{eq:dcf}  
\end{align}

here, B$_\text{pos}$ represents the B-field in the plane of the sky, $Q = 0.5$ is the correction factor recommended by Ostriker et al. ({\it \citealp{ostriker2001density}}), while $Q = 0.28$ is suggested by Liu et al. ({\it \citealp{Liu2022}}) based on their simulation results for cases where the angular dispersion is less than 25$\degree$. In this context, $\rho$ is the gas density, $\sigma_v$ is the nonthermal velocity dispersion, and $\sigma_{\theta}$ denotes the angular dispersion of the plane-of-sky component. In Equation \ref{eq:dcf} ({\it \citealp{crutcher2004drives}}), n(H$_2$) represents the number density (n(H$_2$) = $\rho$/\SI{}{\micro\meter}$_{\text H}$) and $\Delta V_\text{NT}$ represents the FWHM nonthermal velocity dispersion ($\Delta V_\text{NT}$ = $\sigma_\text{v} \sqrt{8ln2}$). We assume a molecular weight $\mu = 2.8$ ({\it \citealp{kirk2013first}}), and this value is applied wherever required. 
C$^{18}$O traces the highly compact, central dense regions of the clumps. To demonstrate this, we estimated the optical depths and column densities of C$^{18}$O. The results indicate that C$^{18}$O is optically thin, making it a reliable tracer of the densest regions within the clumps. These are the same regions probed by the 850 $\mu$m dust emission, allowing us to characterize the level of turbulence at comparable densities.

\subsection*{Measurement of B-field strength using Skalidis $\&$ Tassis method}
Skalidis $\&$ Tassis ({\it \citealp{skalidis2021high}}) highlight the findings from observations indicating that turbulence in the interstellar medium (ISM) exhibits anisotropy and non-Alfv{\'e}nic compressible modes, could have significant relevance. As a result, they put forward a new approach to estimate the B-field strength in the ISM that considers these compressible modes. This can be calculated as,

\begin{align}
    B_{\text{ST}}\, [\mu\text{G}] = \sqrt{2\pi \rho} \, \frac{\sigma_v}{\sqrt{\sigma_\theta}} 
    \approx 1.742 \, \sqrt{n(\mathrm{H}_2)\, [\text{cm}^{-3}]} \, 
    \frac{\Delta V_{\text{NT}}\, [\text{km}\,\text{s}^{-1}]}{\sqrt{\sigma_\theta\, [\text{Degree}]}}
\end{align}

We propose a direct calculation of the B-field strength based on the B-field strength obtained using the DCF method, which can be represented as

\begin{align}
    \hspace{6em} B_\text{ST} = 0.186 \, {\sqrt{\sigma_\theta\, [\text{Degree}]}}\,  B_\text{DCF} \; [\mu\text{G}]
\end{align}

\vskip 6pt
If we have the B-field strength obtained through the DCF method and the angular dispersion ($\sigma_\theta$), it is possible to directly calculate the B-field strength using the ST method. Here we have used Q = 0.5. The B-field strength map obtained by both DCF and ST method are shown in Fig. S\ref{fig:Bfield_lambda}.
While both DCF and ST methods constrain only the plane-of-sky (POS) component of the magnetic field, the line-of-sight (LOS) component remains unconstrained. However a source-specific estimate would indeed be more accurate, such an analysis would require additional observational constraints (e.g., from Zeeman measurements), which are currently unavailable for our target. To account for the missing LOS component, we applied the statistical correction factor, following Crutcher et al. ({\it \citealp{crutcher2004drives}}),
\begin{align}
B_\text{Total} = \frac{4}{\pi} B_\text{POS}.   
\end{align}
This correction accounts for the approximate the total 3D B-field strength from the measured plane-of-sky (POS) component, under the assumption of randomly oriented field segments in a specific source. It is commonly applied in the case of a single source but containing a significantly large number of B-field segments (here in the case of Cep A also, we do not know how each B-field segment is oriented in 3D) and when the LOS component is unconstrained ({\it \citealp{crutcher2004drives,wang2020,pattle2021jcmt,wang2024filamentary}}).

\subsection*{Estimation of Outflow ($E_\mathrm{outflow}$), magnetic ($E_\mathrm{B}$), gravitational ($E_\mathrm{G}$) and turbulent kinetic ($E_\mathrm{K}$) energy}
We followed the procedure described in Section 3.2 of the Xu et al. ({\it \citealp{xu2022}}) to calculate the outflow energy. As we have JCMT/HARP $^{13}$CO emission we have considered this to calculate the outflow energy. This choice was made because $^{13}$CO is generally optically thin and provides a more reliable estimate of the mass where it is detected, avoiding the large optical depth uncertainties associated with $^{12}$CO. From the average spectra of the $^{13}$CO map, we observed that the channels with velocities less than $-$12.5 km/s and greater than $-$8 km/s appear to be influenced by outflows. Therefore, we designated the velocity ranges [-18, -12.5] km/s and [-8, -3] km/s as the blue and red-shifted outflow lobes, respectively. For each channel within these $^{13}$CO outflow lobes, we calculated the energy by considering both the channel mass and channel velocity. To improve accuracy, we subtracted the centroid velocity along each sightline to minimize the impact of large velocity gradients across the entire cloud. We calculated the outflow energy using the formulation ({\it \citealp{xu2022}}),
 \begin{align}
 E_{\mathrm{outflow}} = \frac{1}{2} \sum_{i} M_{\mathrm{i}} ({v}_\mathrm{i} - v)^{2}.
 \end{align} 
 Here M$_i$ is the channel mass, v$_i$ is the channel velocity and v is the centroid velocity. We determined the outflow energies to be (5.73 $\pm$ 0.56) $\times$ 10$^{44}$ and (2.80 $\pm$ 0.30) $\times$ 10$^{44}$ erg for the blue and red-shifted lobes, respectively. It should also be noted that the energies calculated using the $^{13}$CO/HARP data represent lower limits to the outflow energy, since the higher-velocity and more diffuse components can be better traced by $^{12}$CO. The uncertainties in the outflow energies were estimated by propagating the systematic errors arising from the channel noise and mass estimation, together with the statistical uncertainty associated with the centroid velocity. 

To understand the radial distribution of energy in Cep A, we divided the region into subregions based on the radial distance at multiples of 0.5 pc from the center. Within each radial distance, we calculated the magnetic, gravitational, and turbulent kinetic energies. 

The magnetic energy ($E_\mathrm{B}$) is calculated using the relation
\begin{equation}
E_\mathrm{B} = \frac{B^2}{8\pi} \times V,
\end{equation}
where $B$ represents the mean magnetic field strength within the radial distance of $r$, and $V = \frac{4}{3}\pi r^3$ is the corresponding volume. The uncertainty in the $E_\mathrm{B}$ was estimated by propagating the standard error in mean field strength value within the same region.

The gravitational energy ($E_\mathrm{G}$) is determined using
\begin{equation}
E_\mathrm{G} = \frac{3}{5}\frac{G M^2}{r},
\end{equation}
where $M$ is the total mass enclosed within the radial distance $r$.

The turbulent kinetic energy ($E_\mathrm{K}$) is estimated using
\begin{equation}
E_\mathrm{K} = \frac{3}{2} M \sigma_V^2,
\end{equation}
where $\sigma_V$ is average the velocity dispersion measured with respect to the systemic velocity $V_\mathrm{LSR}$ ({\it \citealp{camacho2023kinetic}}), within each radial distance. The uncertainties in the kinetic and gravitational energies are derived from the propagated errors in the mass and velocity dispersion measurements within the region considered.

We note that the estimated energies vary cumulatively with radial distance because their defining physical quantities depend on integrated properties of the cloud. Specifically, gravitational and kinetic energies scale with the total enclosed mass ($M$), while magnetic energy scales with the enclosed volume ($V$). Consequently, as the radial distance increases, both the enclosed mass and volume—and hence the corresponding energies—also increase.


\subsection*{Identification of infalling Gas structures using C$^{18}$O spectral data}
We aim to identify the infall velocity structures within the C$^{18}$O emission that remain unaffected by outflows. We consider channels with velocities ranging from $-$12.5 to $-$8 km/s to achieve this. The other channels with velocities $< -$12.5 and $>-$8 km/s seem to be impacted by the bipolar outflows as they are associated with blue-shifted CO emission and a faint H$_2$ bow shock to the east, as well as with HH 168 to the west ({\it \citealp{cunningham2009pulsed}}). We ensure that the average spectra extracted over the 5$\arcmin$ region of Cep A do not show asymmetric wings. 
Therefore, the channels we have used are not influenced by outflow activity. Since the emission is more concentrated towards the infall velocity structures, we compare the Dendrogram-identified structures to the CMU infall model. To achieve this objective, we adopt the methodology employed by Thieme et al. ({\it \citealp{thieme2022accretion}}), which utilizes the \href{https://dendrograms.readthedocs.io/}{\tt astrodendro} algorithm ({\it \citealp{rosolowsky2008structural}}) to analyze data in position-position-velocity (PPV) space. The dendrogram algorithm initially selects the brightest pixels in the image cube to construct a tree structure. Subsequently, it progressively incorporates fainter pixels until a new local maximum is identified, creating a new structure. These structures are represented as the `leaves' of the tree and are connected by `branches,' which denote pixels that are not local maxima. Eventually, all structures merge to form a complete tree.
Various parameters are provided as input to the algorithm, including a minimum noise threshold for the tree ($\sigma_\text{min}$), a minimum significance level for structures ($\delta_\text{min}$), and the minimum number of pixels required for a structure to be considered independent ($n_{\text{pix,min}}$). We established the values of $\sigma_\text{min}$ as 2$\sigma$, $\delta_\text{min}$ as $\sigma$, and $n_\text{pix,min}$ as 20 pixels. Determining these parameters involved an iterative process of testing different values to identify the configuration that revealed the most cohesive structures. Following this, the algorithm was applied to the C$^{18}$O image cube, identifying a structure highlighted in red within the dendrogram as shown in  Fig. S\ref{fig:dendrogram_structure}. This structure corresponds to an infalling envelope, aligning with the CMU-modeled infalling trajectories, as shown in Figure \ref{fig:cmu_ppv}.

\subsection*{Column density map of the collapsing gas structure}
We produced C$^{18}$O column density under the assumption of local thermal equilibrium (LTE) by employing the procedures mentioned in Mangum \& Shirley et al. ({\it \citealp{mangum2015calculate}}). First, we created the integrated-intensity (moment 0) map of the extracted dendrogram structure, which is a gravitationally collapsing envelope. Then, using the integrted intensity emission $\int S_\nu \Delta \nu $  of each pixel, we made the column density by employing
\begin{align}
    N_\text{C18O} = \frac{3c^2}{16\pi^3\Omega_sS\mu^2\nu^3} \biggl(\frac{Q_\text{rot}}{g_Jg_Kg_I}\biggr) \exp\biggl(\frac{E_{\text u}}{kT_\text{ex}}\biggr)\int S_\nu \Delta \nu.
\end{align}
More details on various parameters can be found at Thieme et al. ({\it \citealp{thieme2022accretion}}). 
The obtained column density of the structure is transformed into the column density of molecular hydrogen (N(H$_2$)) utilizing the abundance ratios: [$^{12}$C/$^{13}$C] = 77, [$^{16}$O/$^{18}$O] = 560, and [H$_2$/$^{12}$CO] = $1.1 \times 10^{-4}$ ({\it \citealp{Frerking1982,eswaraiah2020unveiling}}), resulting in [H$_2$/C$^{18}$O] = $0.61 \times 10^7$. Using the total H$_2$ column density, we then calculated the mass to be 120 $\pm$ 47 M$\odot$.
Similarly, the mass from the moment 0 map of $^{13}$CO can be estimated using the abundance ratio [H$_2$/$^{13}$CO] = 7 × 10$^5$ ({\it \citealp{Frerking1982}}).

\subsection*{Cassen-Moosman-Ulrich (CMU) model}
The CMU model describes the infalling particle trajectories around a central point mass in the gravitational collapse of a rotating, spherically symmetric cloud ({\it \citealp{ulrich1976infall,cassen1981formation,chevalier1983enviroments,terebey1984collapse}}). Key assumptions include solid-body rotation of the cloud at a uniform angular velocity, treating the central star as a point mass, trajectory conservation of specific angular momentum, negligible pressure forces leading to ballistic trajectories, and exclusion of disc, envelope, and B-field.

The path of the infalling matter described by parabolic trajectory and the angular relations is given:
\begin{align}
            r &= \frac{\sin^2\theta_0} {1-\frac{\cos\theta}{\cos\theta_0}} r_c,\nonumber\\
    \cos\theta &= \cos\nu \cos\theta_0, \text{and} \nonumber\\
       \tan\nu &= \tan(\phi - \phi_0) \sin\theta_{0}.
\end{align}
Here, $\theta_{0}$ is the polar angle of the orbital plane for the particle
trajectory, $\phi_{0}$ is the azimuthal angle of the orbital apastron, and $\theta$
and $\phi$ are the polar and azimuthal angles of the particle after a time `$t$.' $r_c$ is the centrifugal radius, and $\nu$ denotes the angle of direction for the particle, measured from the origin to its farthest point from the star. 

We fixed the inclination angle (62$\degree$) and position angle (44$\degree$) based on previous observational results ({\it \citealp{patel2005}}). We modeled the trajectory with $\theta_0$ varying from 1 to 180$\degree$ and $\phi_0$ from 0 to 360$ \degree$ with a protostellar mass of 15 M$_{\odot}$. To finalize the framework, we also incorporated the velocity of the particles to compare in the PPV space. For more details, see Thieme et al. ({\it \citealp{thieme2022accretion}}).

In the CMU model, trajectories trace the infalling matter uniformly from all directions, thereby depicting the scenario of global gravitational collapse. 
We have compared the observed dendrogram structure to the CMU-modeled trajectories in the PPV space by matching every model data point lying spatially within the half pixel-width ($\sim$7.27$\arcsec$) of RA and Dec of the dendrogram data point and also lying within the half of spectral resolution ($\sim$ 0.17 km/s) of C$^{18}$O cube. We determined the infall velocity, time and rate using the equation outlined in Pineda et al. ({\it \citealp{pineda2020protostellar}}) as,
\begin{align*}
    v_\text{ff}= -\sqrt{\frac{2GM}{R}}\\
    t_\text{ff} = \sqrt{\frac{R^3}{GM}}\\
    \dot{M}_\text{envelope} = \frac{M_\text{envelope}}{t_\text{ff}}
\end{align*}

\clearpage

\clearpage

\section*{Acknowledgements}

We thank Marco Padovani for the insightful discussions regarding the provision of 3D B-field modeling. We also thank Yuehui Ma for the help with determining the optical depths and column densities of molecular lines data. We thank Nathaniel Cunningham and John Bally for supplying the 2.12 \SI{}{\micro\meter} H$_2$ continuum-subtracted data and Henrik Beuther for providing the SMA polarimetry data. We thank Ian Stephens, M. S. Nanda Kumar, Jihye Hwang,  and Sheng-Jun Lin for the helpful discussions.

This work is partially funded by NSFC grant No. 12588202 and new Cornerstone Science Foundation.
These observations were obtained by the James Clerk Maxwell Telescope, operated by the East Asian Observatory on behalf of The National Astronomical Observatory of Japan; Academia Sinica Institute of Astronomy and Astrophysics; the Korea Astronomy and Space Science Institute; the National Astronomical Research Institute of Thailand; Center for Astronomical Mega-Science (as well as the National Key R\&D Program of China with No. 2017YFA0402700). Additional funding support is provided by the Science and Technology Facilities Council of the United Kingdom and participating universities and organizations in the United Kingdom and Canada. This research  utilized data from the Milky Way Imaging Scroll Painting (MWISP) project, which is a multi-line survey in $^{12}$CO/$^{13}$CO/C$^{18}$O along the northern galactic plane with PMO-13.7m telescope. We are grateful to all the members of the MWISP working group, particularly the staff members at PMO-13.7m telescope, for their long-term support. MWISP was sponsored by National Key R\&D Program of China with grant 2017YFA0402701 and by CAS Key Research Program of Frontier Sciences with grant QYZDJ-SSW-SLH047. 



\subsection*{Funding}

\begin{itemize}[noitemsep, nolistsep, label={}]

\item Ramanujan Fellowship grant by Science and Engineering Research Board (SERB) No. RJF/2020/ 000071 (CE). 
\item Core Research Grant (CRG; sanction order number CRG/ 2023/008710) awarded by the Anusandhan National Research Foundation (ANRF) under Science and Engineering Research Board (SERB), Govt. of India (CE). 
\item DST-INSPIRE Fellowship from the Department of Science and Technology (DST) No. IF200294 (PS). 
\item National Science and Technology Council with the grant No. NSTC 114-2112-M-007-001 (SPL).
\item Department of Atomic Energy, Government of India, under project identification No. RTI 4002 (DKO). 
\end{itemize}

\subsection*{Author Contributions}

\begin{itemize}[noitemsep, nolistsep, label={}]
\item PS: Writing – original draft, Conceptualization, Investigation, Writing – review \& editing, Methodology, Resources, Data curation, Validation, Supervision, Formal analysis, Software, Project administration, Visualization.

\item CE: Writing – original draft, Conceptualization, Investigation, Writing – review \& editing, Methodology, Resources, Funding acquisition, Validation, Supervision, Project administration, Visualization
\item DL: Conceptualization, Writing – review \& editing, Methodology, Resources, Funding acquisition
\item EVS: Conceptualization, Writing - review \& editing, Methodology, Supervision
\item GCG: Writing – original draft, Conceptualization, Investigation, Writing – review \& editing, Methodology, Formal analysis
\item TJT: Conceptualization, Writing – review \& editing, Methodology
\item MRS:  Writing – original draft, Writing – review \& editing, Visualization
\item JW: Conceptualization, Investigation, Writing – review \& editing, Methodology
\item SPL: Conceptualization, Investigation, Writing - review \& editing, Methodology, Resources, Validation
\item WPC: Conceptualization, Writing – review \& editing, Resources, Validation, Supervision
\item DKO: Writing – review \& editing, Resources, Validation, Supervision, Visualization
\end{itemize}

\subsection*{Competing Interests}
The authors declare no competing interests.

\subsection*{Data, Code, and Materials Availability}
All data and code needed to evaluate and reproduce the results in this paper are present in the paper and/or the Supplementary Materials. The data for this study have been deposited in the Dryad Repository (\url{https://doi.org/10.5061/dryad.sqv9s4ngq}). This study did not generate new materials.

\clearpage

\begin{center}  
Supplementary Materials for
\vskip 2pt
{\bf Evidence for the gravity-driven and
magnetically-regularized gas flows feeding the
massive protostellar cluster in Cep A}
\vskip 2pt
Panigrahy Sandhyarani, Chakali Eswaraiah {\it et. al}
\vskip 2pt
$^*$Corresponding author. Email: eswaraiahc@iisermohali.ac.in
\end{center}
{\bf This PDF file includes:}

\vspace{.2 cm}
\hspace{.5cm} Figs. S1 to S5

\clearpage
\setcounter{figure}{0}
\renewcommand{\figurename}{ Fig. S}


\begin{figure*}[hbt!]
\centering
\includegraphics[width=1\textwidth]{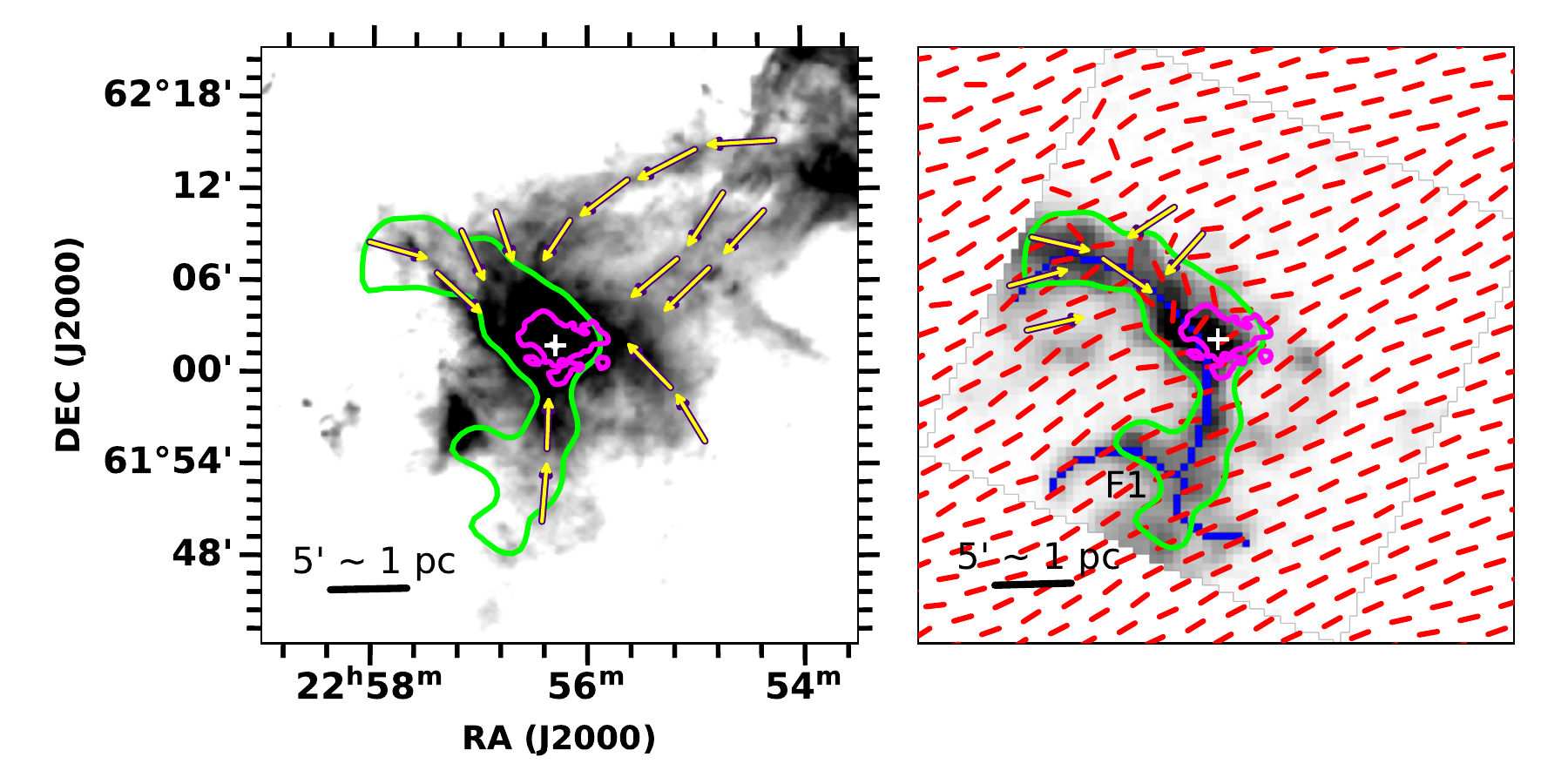}
\caption{{\bf Large-scale filamentary cloud and the signature of bend in B-fields.} Left panel: The {\it Herschel}/SPIRE 250 \SI{}{\micro\meter} dust continuum map revealing the large-scale extended filaments, as identified using yellow arrows, in Cep A cloud. Right panel: The cloud-scale B-field geometry (red segments) using the low-resolution (5$\arcmin$) {\it Planck} 850 \SI{}{\micro\meter} dust continuum polarization data. These B-fields are overlaid on the moment 0 map, revealing gas structure within the channels of $-12.5$~--~$-11$ km/s. In both panels, a green contour, drawn at $^{13}$CO integrated intensity of 3.5 K km/s, traces the identified $^{13}$CO gas structure that is identified as F1 (see Fig. S\ref{fig:LARGE_SCALE_SMALL_SCALE}). The blue curve delineates the filament identified by the FILFINDER algorithm. Yellow vectors guide the bend in the B-field, possibly driven by the cloud's gravity. `+' sign represents the HW2 protostar. Magenta contour represents the POL-2 dust continuum emission (Stokes I) at 32 mJy/beam (equivalent to 10$\sigma$, where $\sigma$ = 3.2mJy/beam).}
\label{fig:planck_on_herschel_pmo}
\end{figure*}

\begin{figure*}
\centering
\begin{picture}(0,0)  
   {\small PMO $^{13}$CO (2 ~--~ 1) data} 
\end{picture}
\includegraphics[width=17
 cm, height=5 cm] {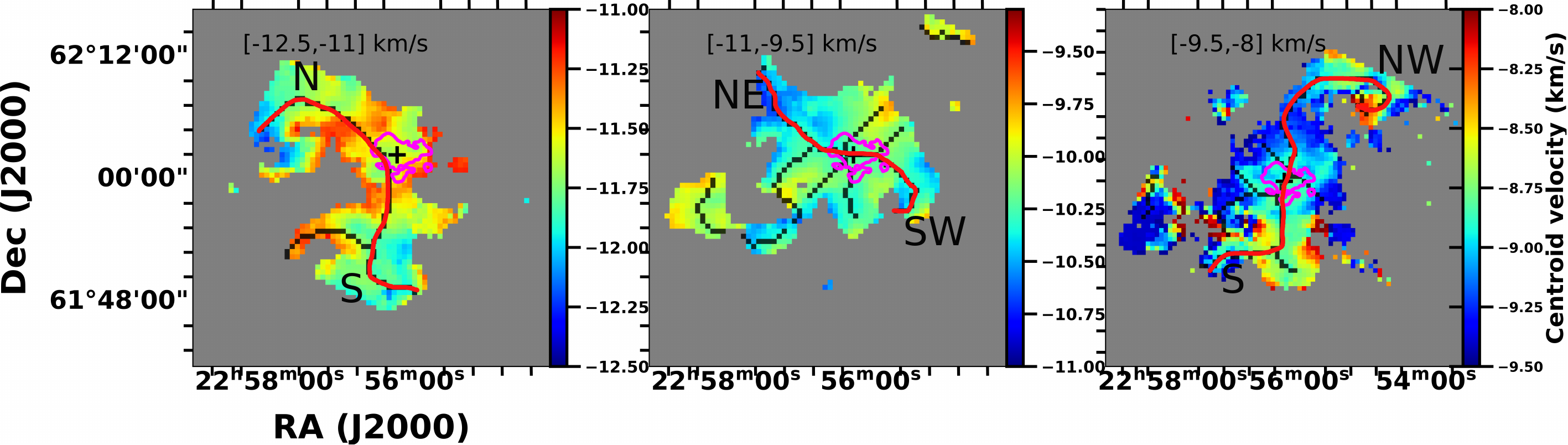}
\includegraphics[width=17
 cm, height=5 cm] {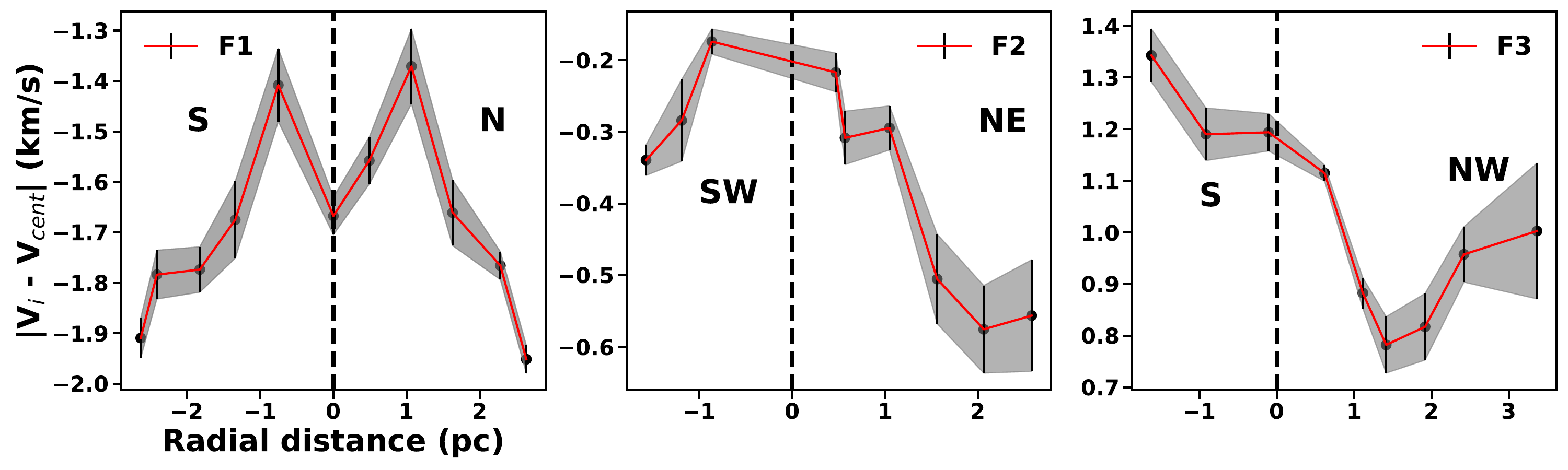}
\begin{picture}(0,0)  
   \put(-40,5){\small JCMT/HARP C$^{18}$O (3 ~--~ 2) data} 
\end{picture}
\includegraphics[width=17
 cm, height=7 cm] {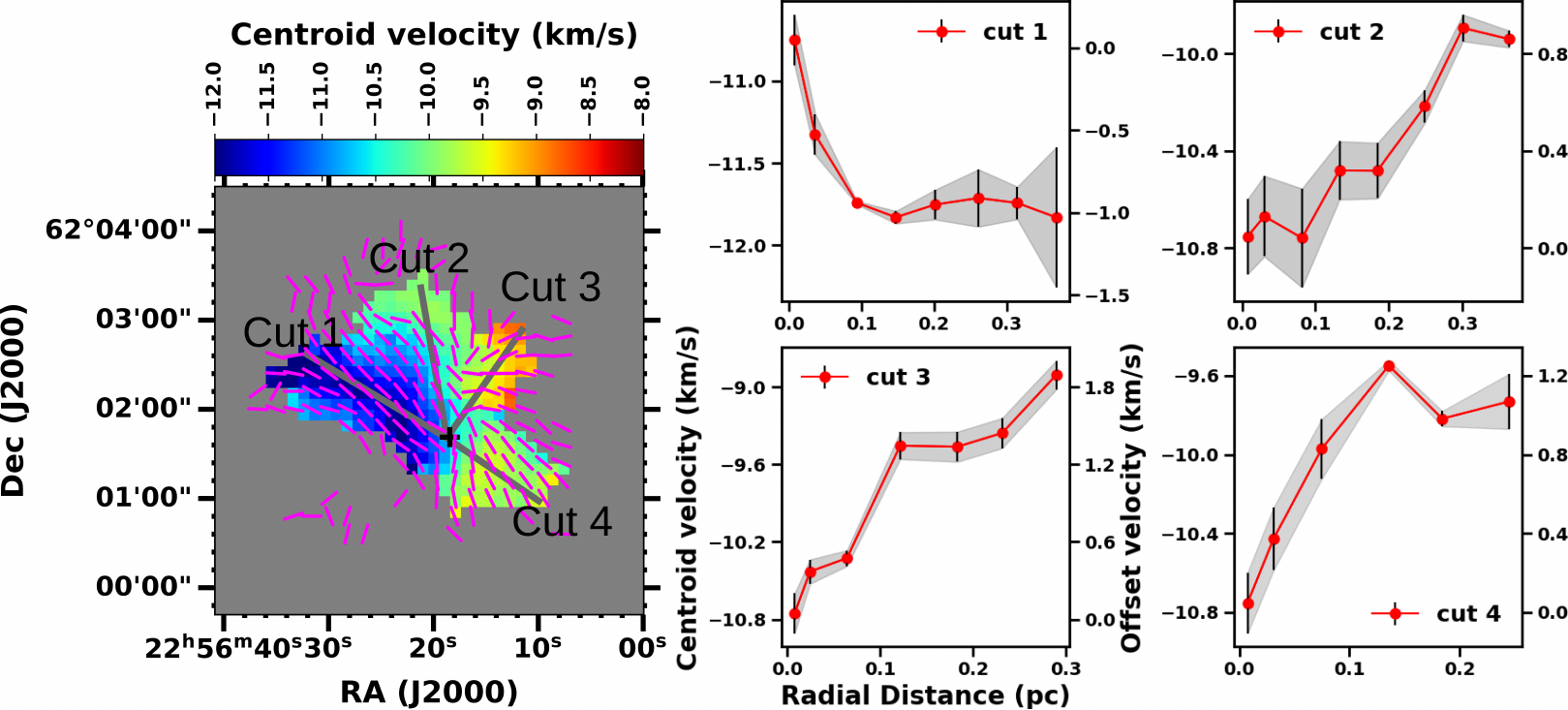}
\caption{{\bf Gas kinematics revealed by PMO and HARP molecular line data.} Gas structures and velocity gradients at the cloud and clump scales inferred based on the moment 1 maps using PMO $^{13}$CO (top two rows) and JCMT/HARP C$^{18}$O (bottom panels) data. The magenta contour in the top panel corresponds to POL-2 dust continuum emission (Stokes I) at 32 mJy/beam (equivalent to 10$\sigma$, where $\sigma = 3.2$ mJy/beam is the rms noise in the dust continuum map. The main filaments are highlighted in red, along which the velocity gradients are taken, is shown in the middle panel. The `+’ sign represents the position of the HW2 protostar. The bottom panel displays four distinct cuts on the JCMT/HARP C$^{18}$O moment 1 map, while the bottom right panels present the corresponding velocity gradients extracted along these cuts. The dashed vertical line in the 2nd row and zero pc in the bottom right panels mark the position of the HW2 protostar.}
\label{fig:LARGE_SCALE_SMALL_SCALE}
\end{figure*}

\begin{figure*}
\centering
\includegraphics[width=7.5cm, height=5.6cm]{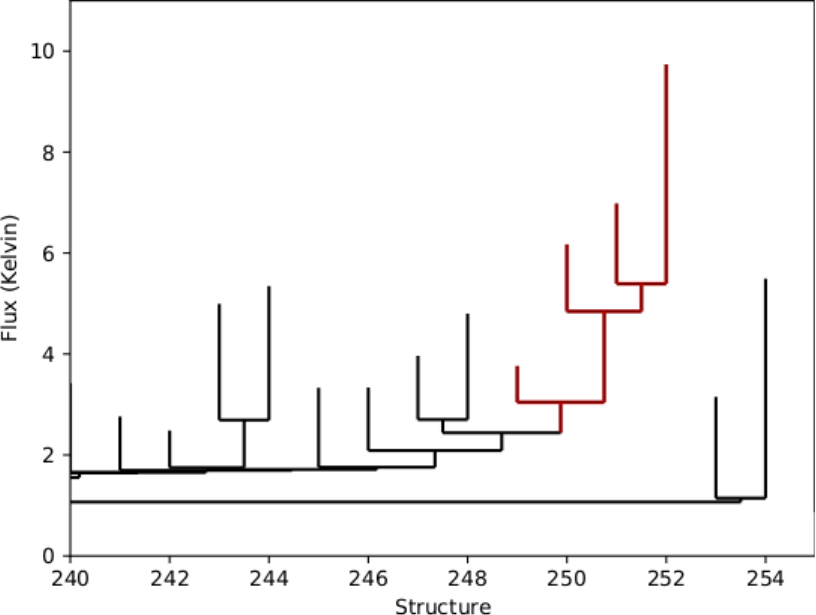}
\hspace{0.3cm}
\includegraphics[width=7.5cm, height=5.6cm]{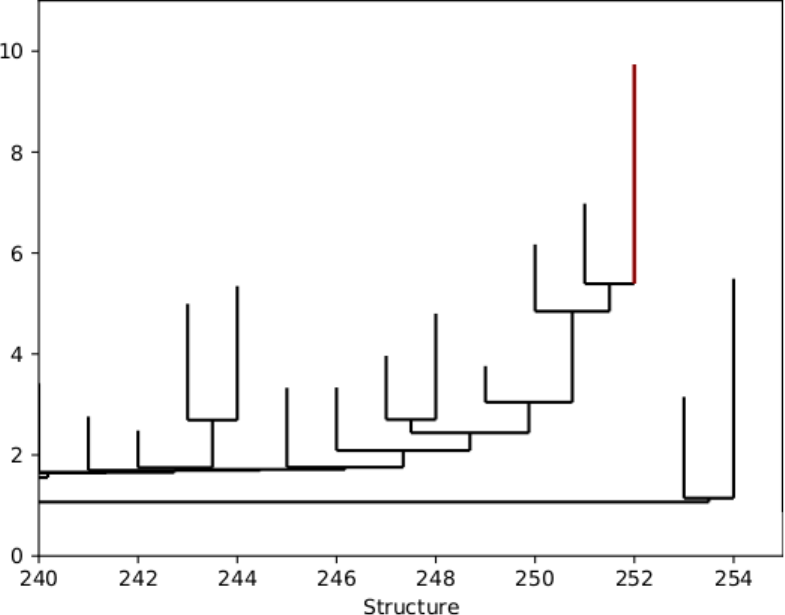}
\includegraphics[width=7.5cm, height=6cm]{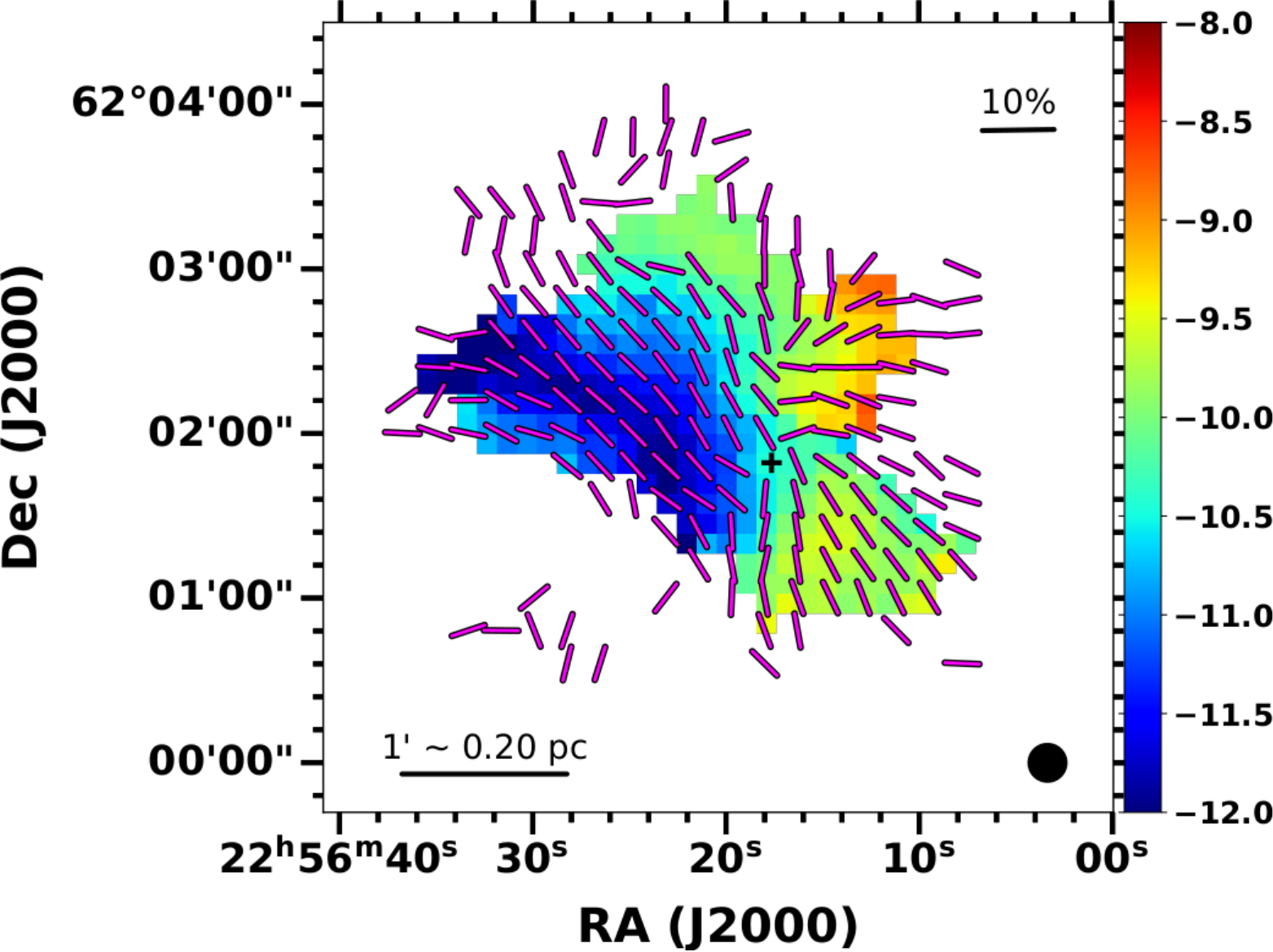}
\includegraphics[width=7.5cm, height=6cm]{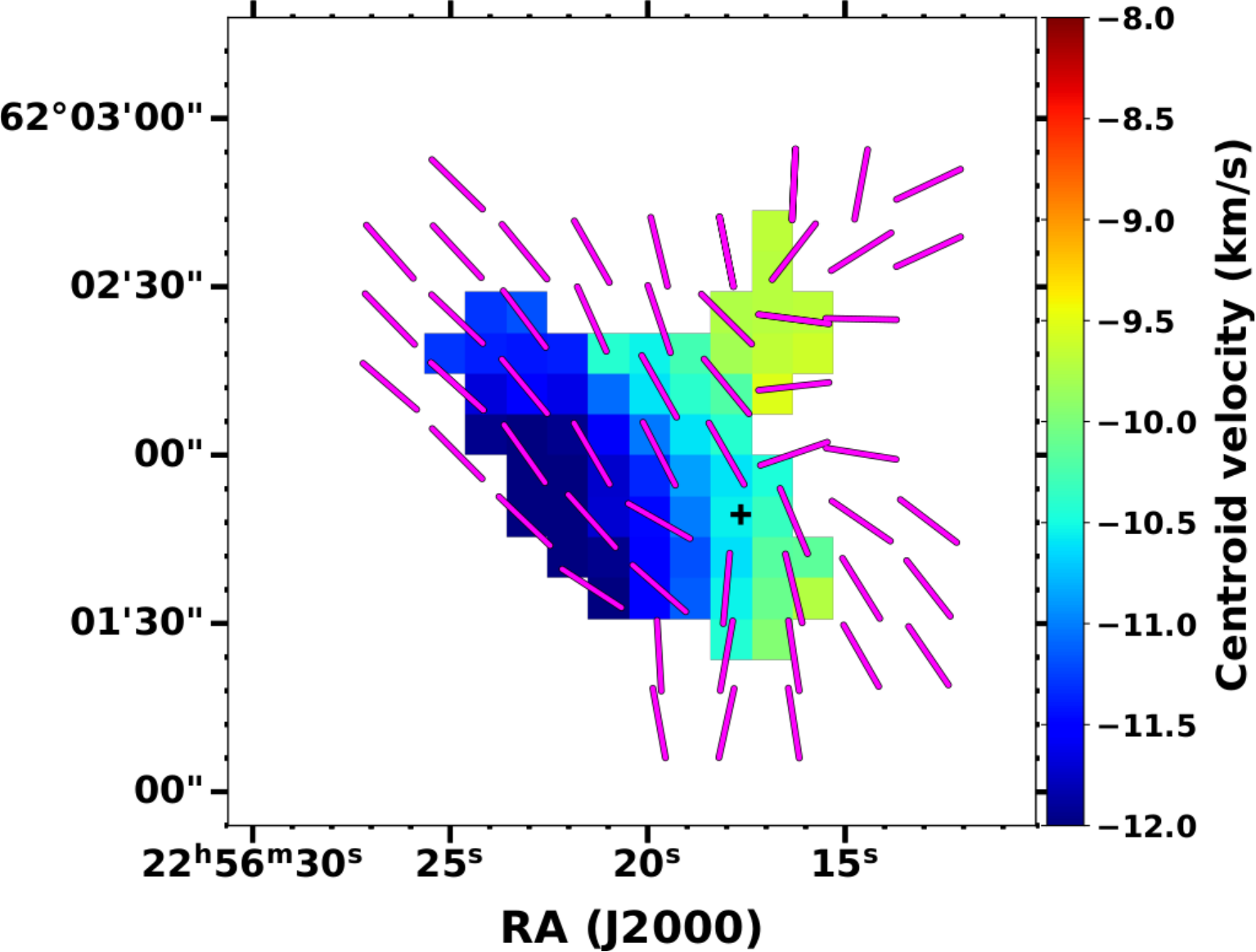}
\caption{{\bf Collapsing gas structures identified based on the Dendrogram analyses.} (Left): Dendrogram tree of C$^{18}$O cube. Highlighted in red is the gravitationally collapsing envelope. (Right): Same as a left panel but showing the more prominent leaf of the dendrogram tree. (Bottom panel) Moment 1 map of the identified gas structures based on the dendrogram analyses depicted in the left and right panel. An infalling envelope comprises three gas structures that exhibit different velocities~--~visible in blue, cyan, and yellow/red, as shown in the left and right panels. Conspicuously, the B-field orientation associated with each gas structure is also different. The clump-scale infalling gas structure is extended close to the massive protostar Cep A HW2. The B-field got twisted at the location where these gas structures merge, as shown on the right panel.}
\label{fig:dendrogram_structure}
\end{figure*}

\begin{figure*}
\centering
\includegraphics[width=8cm, height=5.6cm]{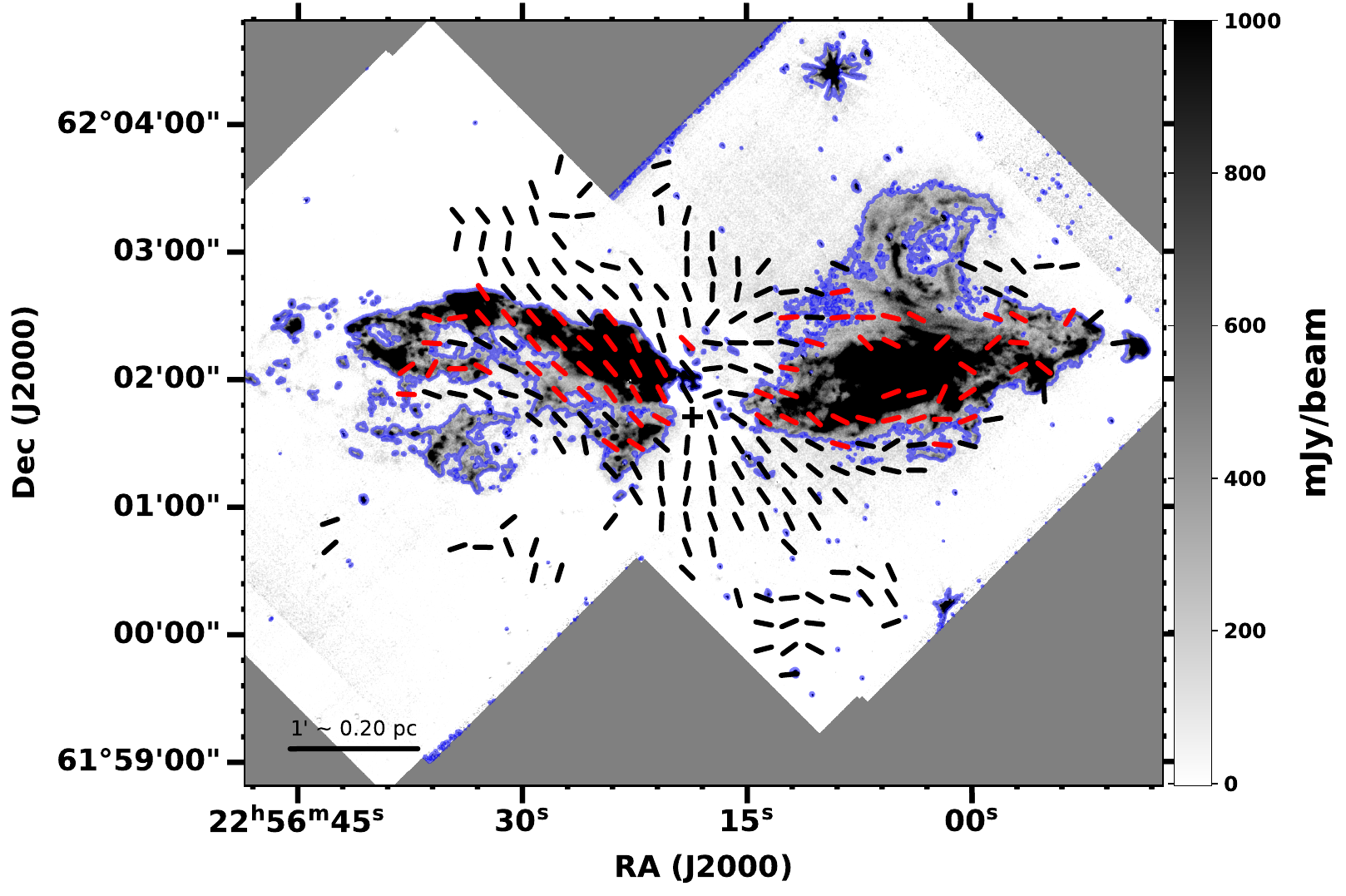}
\includegraphics[width=7cm, height=5.6cm]{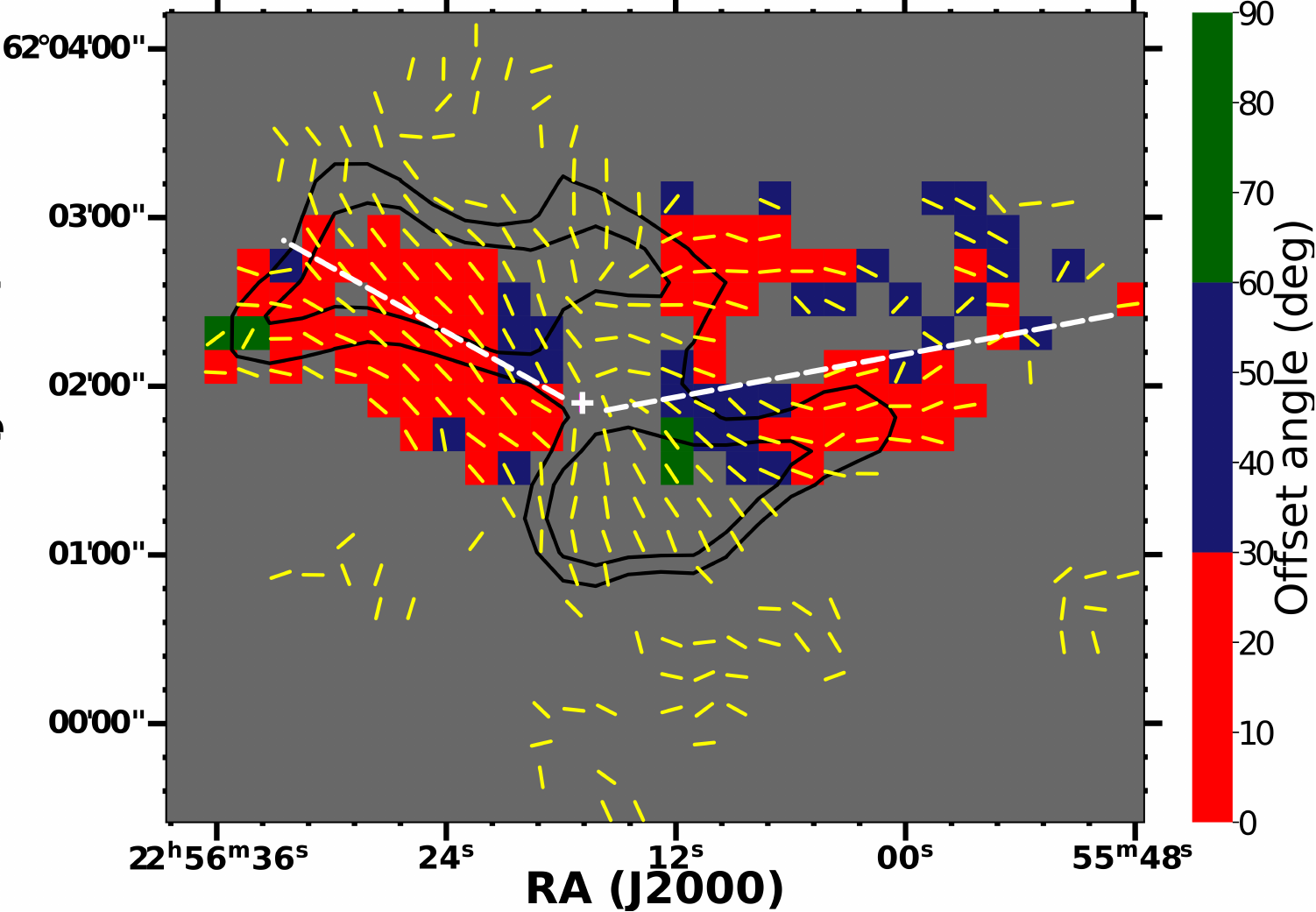}
\caption{{\bf The orientation of B-fields and outflow in Cep A.}
(Left )The shock-excited H$_\mathrm{2}$ emission at 2.12 \SI{}{\micro\meter} tracing bipolar outflows emanated from the Cep A HW2 ({\it \citealp{cunningham2009pulsed}}).
Blue contour is drawn at 200 mJy/beam. Red segments are distributed within the H$_{2}$ emission of 200 mJy/beam. The black segments lie in the regions with negligible H$_{2}$ emission of $<$ 200 mJy/beam may trace the B-fields unperturbed by the outflows. (Right) Offset map between the position angles of outflows and B-field. White dashed lines represent the mean position angles of the outflows.} 
\label{fig:outflows_and_Bfieldmorphology}
\end{figure*}

\clearpage

\begin{figure*}
\centering
\includegraphics[width=8cm, height=6.5cm]{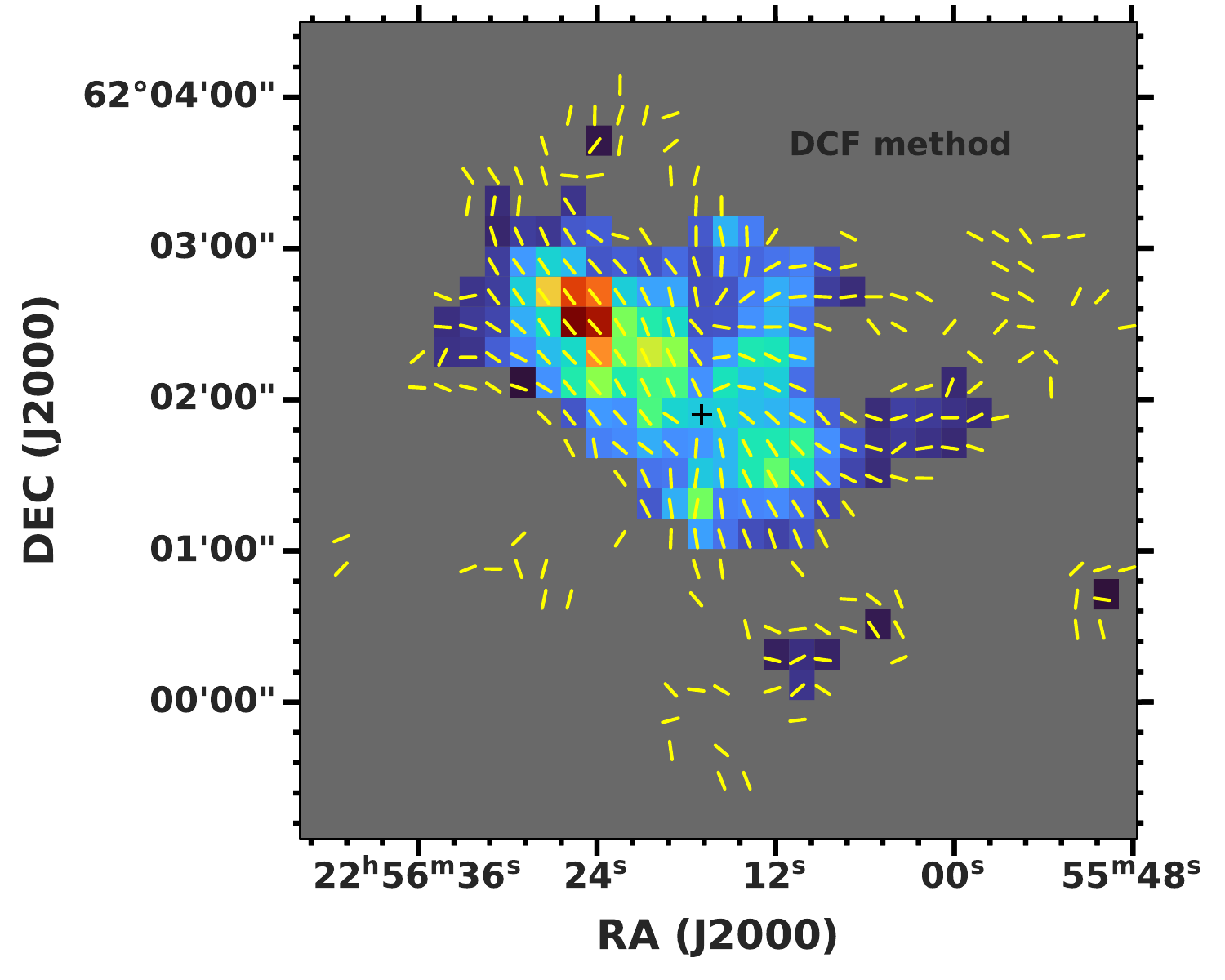}
\hspace{0.3cm}
\includegraphics[width=7cm, height=6.5cm]{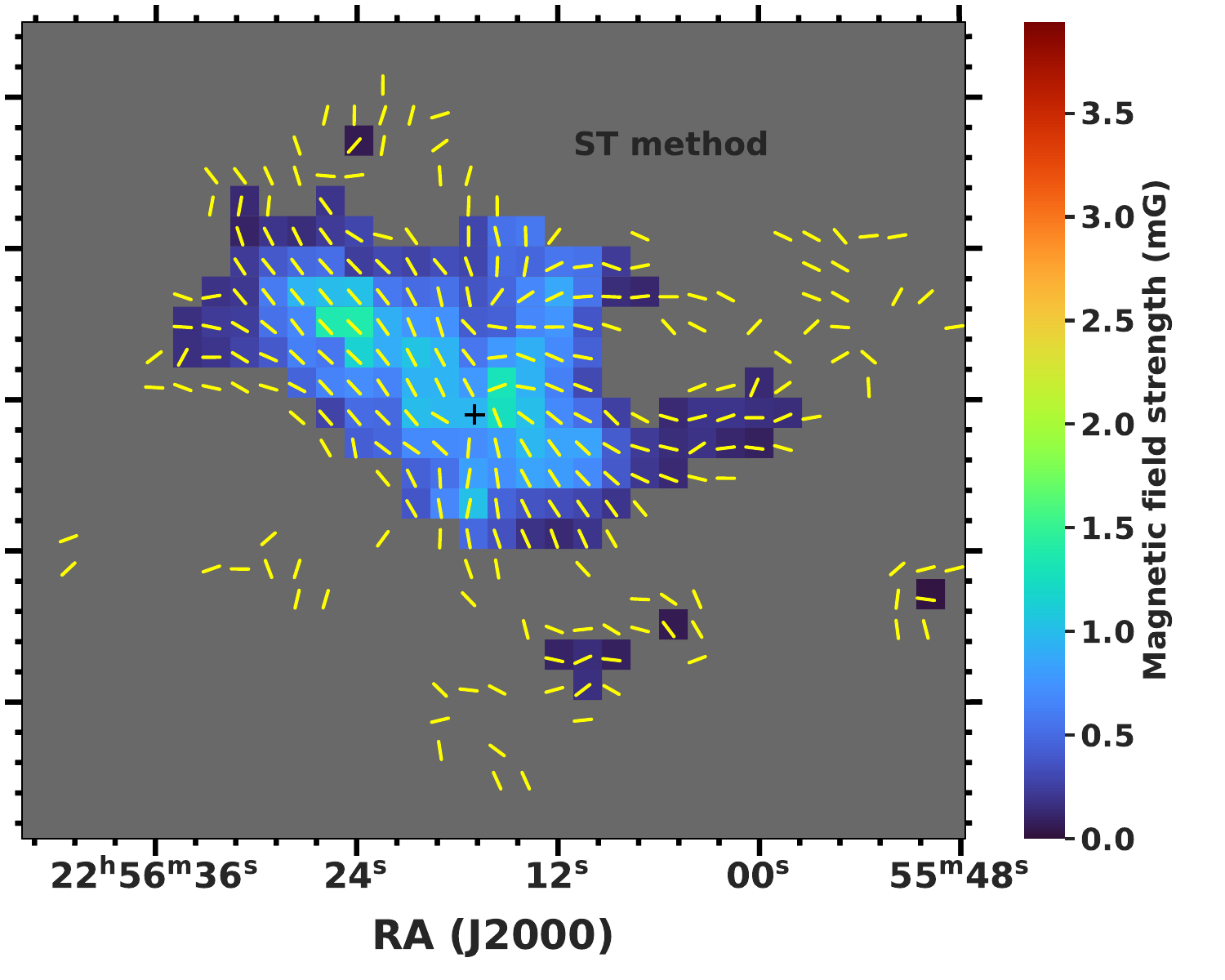}
\caption{{\bf B-field strength map.} B-field strength map using the DCF method (left) and the ST method (right). Yellow segments represent the POL-2 B-field. The `+' sign represents the position of the HW2 protostar. The colorbar applies to both the maps produced using the ST and DCF methods.}
\label{fig:Bfield_lambda}
\end{figure*}

\end{document}